# Security of Healthcare Data Using Blockchains: A Survey

*Mayank Pandey[1*], Rachit Agarwal[2], Sandeep K. Shukla[2], Nishchal K. Verma[1]*

[1] Department of Electrical Engineering, Indian Institute of Technology Kanpur, U.P., India
[2] Department of CSE, Indian Institute of Technology Kanpur, U.P., India
{pandeym, nishchal}@iitk.ac.in, {rachitag, sandeeps}@cse.iitk.ac.in

* Corresponding author

*Abstract:* The advancement in the healthcare sector is entering into a new era in the form of Health 4.0. The integration of innovative technologies like Cyber-Physical Systems (CPS), Big Data, Cloud Computing, Machine Learning, and Blockchain with Healthcare services has led to improved performance and efficiency through data-based learning and interconnection of systems. On the other hand, it has also increased complexities and has brought its own share of vulnerabilities due to the heavy influx, sharing, and storage of healthcare data. The protection of the same from cyber-attacks along with privacy preservation through authenticated access is one of the significant challenges for the healthcare sector. For this purpose, the use of blockchain-based networks can lead to a considerable reduction in the vulnerabilities of the healthcare systems and secure their data. This chapter explores blockchain's role in strengthening healthcare data security by answering the questions related to what data use, when we need, why we need, who needs, and how state-of-the-art techniques use blockchains to secure healthcare data. As a case study, we also explore and analyze the state-of-the-art implementations for blockchain in healthcare data security for the COVID-19 pandemic. In order to provide a path to future research directions, we identify and discuss the technical limitations and regulatory challenges associated with blockchain-based healthcare data security implementation.

## 1  Introduction

With the advancements in technology, various tools and methodologies exist that facilitate the healthcare sector's efficient and effective functioning. The influx of large volumes of data coming from healthcare-related Internet of Things (IoT) based devices further catalyzes the functioning. For example, IoT devices provide a way to transform hand-written diagnostic and other related reports into digital reports. They also provide a way to make the storage and sharing of such reports convenient with different stakeholders (such as patients, doctors, pharmaceutical suppliers, health insurance people, and healthcare administrators). On top of this, analytics on healthcare-related data paves the way for many related services to co-exist. For example, in the case of a pandemic like COVID-19, the vaccine's development is based on the available genetic data (Le et al., 2020), while the use of Remdesivir for treatment is based on clinical trial data (Beigel et al., 2020). Further, as another example, in (Brown et al., 2020), the authors propose using smartphones to capture and analyze cough sounds for identifying potential infection with COVID-19. Apart from this, analysis of the day-to-day healthcare data provides policymakers necessary inputs to make policies.

A significant healthcare data component comprises the Electronic Healthcare Records (EHR) (Stafford & Treiblmaier, 2020). EHR contains private medical diagnosis information of a patient. Apart from EHR, there are various other types of healthcare data. These include: (a) personal health record (PHR) (Bietz et al., 2016) that contains the data such as the physiological health parameters and allergies information of the patient, (b) pharmaceutical/medicine data that includes information on the manufactured medications and its clinical trial data (Burbidge et al., 2001), (c) health insurance data that comprises of information related to the insurance policy of the patient, payment information and data on the availed insurance-related services (Moulis et al., 2015), and (d) data generated and used for research in healthcare such as those of cough sounds for COVID-19 detection(Brown et al., 2020) and X-ray images for pneumonia classification (Stephen et al., 2019). Such immense data help doctors gain more understanding about their patients and provide better healthcare services to them. Nonetheless, different stakeholders exploit different healthcare data for (a) enhancing their services, such as pharmaceutical representatives influencing doctors (Fugh-Berman & Ahari, 2007) or (b) doing illegal activities such as drug abuse (Berenson & Rahman, 2011).

A major challenge with such healthcare data is to ensure its security from various cyber-attacks such as unauthorized access and tampering. In (Kumar & Walker, 2017), the authors provide in-depth analysis of: (a) what threats exist (such as EHR in hospitals are not being well protected) for the healthcare data, (b) how many people and organizations providing healthcare services are affected by the breach (such as Athena group hack), and (c) the behavior responsible for the success of such attacks (such as the habit of personal Internet surfing and social media access on workplace computers along with low cybersecurity awareness). Security of the healthcare data has recently been the focus of many policymakers due to various cyber-attack incidents, such as the one on the University of Vermont (UVM) Medical Centre and the Ryuk ransomware attack in October 2020. In (Dyrda, 2020), the author provides an analysis of the significant cyber-attacks that happened in 2020 on healthcare data and reiterates the necessity of data security education and awareness among the people working in the healthcare sector.

As the threats to such data increase, preventive measures and research to improve healthcare data security are highly critical. One viable solution for improving healthcare data security is to use Blockchain technology, which ensures the integrity, immutability, and traceability of the data. Blockchain technology was first introduced in 2008 (Nakamato, 2008) with the introduction of Bitcoin as an alternative to the existing centralized banking and payment structures. Here, the author presented blockchain as a decentralized ledger with the capability of storing the transaction records in blocks that are serially and sequentially connected. Blockchain operates as a decentralized peer-to-peer network, where the participants (or the nodes in the network) carry out transactions, transaction verification, and mining, and willing participants have a copy of the blockchain stored with them, providing redundancy.

In healthcare, applications of blockchain technology are in the field of data management and sharing (Hölbl et al., 2018), pharmaceutical supply chain management (Khezr et al., 2019), and secure medical data storage; and log management (De Aguiar et al., 2020). In (Shi et al., 2020), the authors survey and discover the research opportunities regarding the collaboration of blockchain technology with other emerging technologies such as big data, machine learning, and IoT. In (Tariq et al., 2020), the authors examine blockchain technology's role in the data security of resource-constrained medical IoT devices. The authors identify that such devices are vulnerable to attacks such as forgery and data tampering. They also survey blockchain-based solutions that address the security issues faced by the healthcare systems. In (Hardin & Kotz, 2019), the authors provide an analysis of the challenges faced during the implementation of blockchain technology in healthcare systems in terms of security parameters such as integrity, confidentiality, access control, and interoperability. The authors identify that such technology benefits for healthcare security come with their own sets of challenges, such as deciding on data requirements and individual privacy. In (Radanović & Likić, 2018), the authors explore the opportunities for blockchain technology in medicine and argue that such technology can mitigate the shortcomings related to different types of data such as EHR. They also provide a discussion on the benefits of blockchain technology not only with respect to data security but also with respect to other factors such as processing time reduction, cost reduction, and transparency. Note that, henceforth, in this chapter, we refer to blockchain technology as blockchain.

In terms of healthcare data security, all the aforementioned work focuses on the solutions for securing EHR data. Pharmaceuticals, administration, and health insurance are also useful as the organizations relevant to these components are part of the healthcare domain and are actively involved in sharing, accessing, and using the generated data. Healthcare data security also encompasses protecting information such as medical diagnosis data, health insurance data, pharma supply chain data, and biomedical research data.

In this chapter, we assess the role of the blockchain in strengthening healthcare data security through the fundamental questions: why, what, who, when, and how, i.e., 4W and 1H. More specifically, we answer why we need a blockchain for the security of healthcare data? What (Which type) are the healthcare data areas we need a blockchain for implementing? Who (Which organizations and individuals) need a blockchain for protecting their healthcare data? When do we need a blockchain for the security of healthcare data, and how to implement a blockchain for the same?

Based on the above questions, our key contributions through this chapter are:
- **Survey**: Using the 4W1H methodology, we present an understanding on what data, when we need, why we need, who needs, and how state-of-the-art techniques use blockchains to secure healthcare data. We also identify the gaps present in the state-of-the-art methods related to the implementation of blockchain in healthcare data from the security perspective and survey them. We also provide a survey of technical and regulatory challenges faced in the process of implementation of blockchain-based healthcare data security systems. We identify that a proper user interface for the blockchain-based systems is warranted. Also, the immutability of data on the blockchain is in direct contention with the right to be forgotten, which comes under the right to privacy.
- **Application-Specific**: We explore blockchain's role towards healthcare data security in pandemic situations like COVID-19.
- **Research-Perspectives**: Based on the survey, we provide prospective future research and development directions that can benefit authorities and researchers in the field.

The remaining chapter is organized as follows. First, via section 2, we provide an understanding of the healthcare data types and their managing institutions. Then, in section 3, we provide services that are rendered using such healthcare data. In section 4, we discuss and explain the different security issues present when using conventional healthcare data storage and management systems. In section 5 and section 6, we provide the security parameters required for healthcare data protection and an overview of healthcare-relevant blockchain functionalities and advantages, respectively. In section 7, we discuss the functionality of different state-of-the-art blockchain-based solutions that enable data security. In sections 8 and 9, we present various technical limitations and regulatory challenges that exist while implementing a blockchain-based solution for healthcare data security, respectively. In section 10, as a case study, we study how blockchains can help in achieving data security in the case of a pandemic such as COVID-19. In the end, in section 11, we present the conclusion and perspective to future research directions.

## 2 Types of healthcare data systems and managing institutions

The healthcare data is a pool of all the information relevant to the organizations and people associated with the healthcare sector. Before we discuss blockchain implementation for healthcare data security, we provide the details of the healthcare sector data. This section provides a brief discussion on the different categories of data associated with the healthcare sector and their managing institutions. In the healthcare sector, the data systems operate through the functionalities comprising the generation, access control, and individual health data storage.

We broadly categorize the healthcare data into three categories: personal data, pharmaceutical data, and insurance-based data. We further classify personal data into patient data and healthcare professional data. Among the patient data, the most common and frequently generated patient records are the **Electronic Health Records** (EHRs) (Hayrinen et al., 2008). An EHR contains personal details, physiological health parameters, medical history, laboratory-generated medical test results, and pharmaceutical prescription data of a patient. Healthcare institutions such as hospitals and clinics generate EHR data based on the healthcare professionals' and laboratories' diagnoses. Specialist third-party companies and vendors generally do the storage and management of the EHRs, while the healthcare institutions and professionals are their clients (Mandl & Kohane, 2012).

Other than EHRs, **Personal Health Records** (PHRs) are the records generated and owned by the patients (Win et al., 2006) where each patient provides access to his PHR information based on the need. A PHR is usually generated using smartphones and various other IoT-based wearable medical devices. It contains information on the general well-being of a patient, including data about blood pressure, heart rate, body temperature, allergies, vaccination history, and previous surgeries. PHRs pave the way for remote assistance and decrease the response time in case of an emergency. The healthcare professionals render their advice to the patients based on the data available in both EHR and PHR. PHR data is managed by the individuals themselves with the help of specialized third-party tools. For example, the use of smartphone-based applications, web-based storage services (such as HealthVault (Sunyaev et al., 2010)), and specialized software for data storage (Carrion et al., 2011).

In the personal data category, another type of data is the information about the **healthcare professionals, administrative staff, and researchers** (Fugh-Berman & Ahari, 2007). This data contains personal details, professional qualifications, work timings, and the details of their professional tasks. The data encompassing these details is vital for the smooth day-to-day operations of healthcare organizations. The respective institutions to which the person is affiliated manage the storage and management of such personal information.

Besides the personal data, the **data on pharmaceuticals** is also vital. The pharmaceutical data have different attributes, such as those related to clinical trial data, medication manufacturing information, and pharma supply chain data containing the distribution information. The clinical trial data is essential for the invention of new treatment methods in the form of medications and vaccines. The medication manufacturing information includes the medication's composition, information related to side effects and allergies, and clinical trial data for research and development activities. In addition, data on manufacturing date, expiry date, and dosage information for manufactured medications is also critical. The information on medications for various ailments is essential for doctors to proceed with treatment. On the other hand, the supply chain is also a crucial part of the pharmaceutical industry. The supply chain data is essential in the distribution of raw material and the final product. It constitutes information such as the list of distributors, quantity, distribution, product storage, and transportation-related condition information. Such pharmaceutical data is stored and managed by the pharmaceutical companies either by themselves or through specialist third-party organizations.

For smooth processing of patient treatment, **health insurance information and the related data** are also essential (Pitacco, 2014). Such data is usually shared between insurance companies and hospitals. This data contains both health and financial information of a patient along with the information on availed insurance facilities. More specifically, it includes details such as personal details, medical history, and insurance plan preference. Such data is stored and managed by insurance companies.

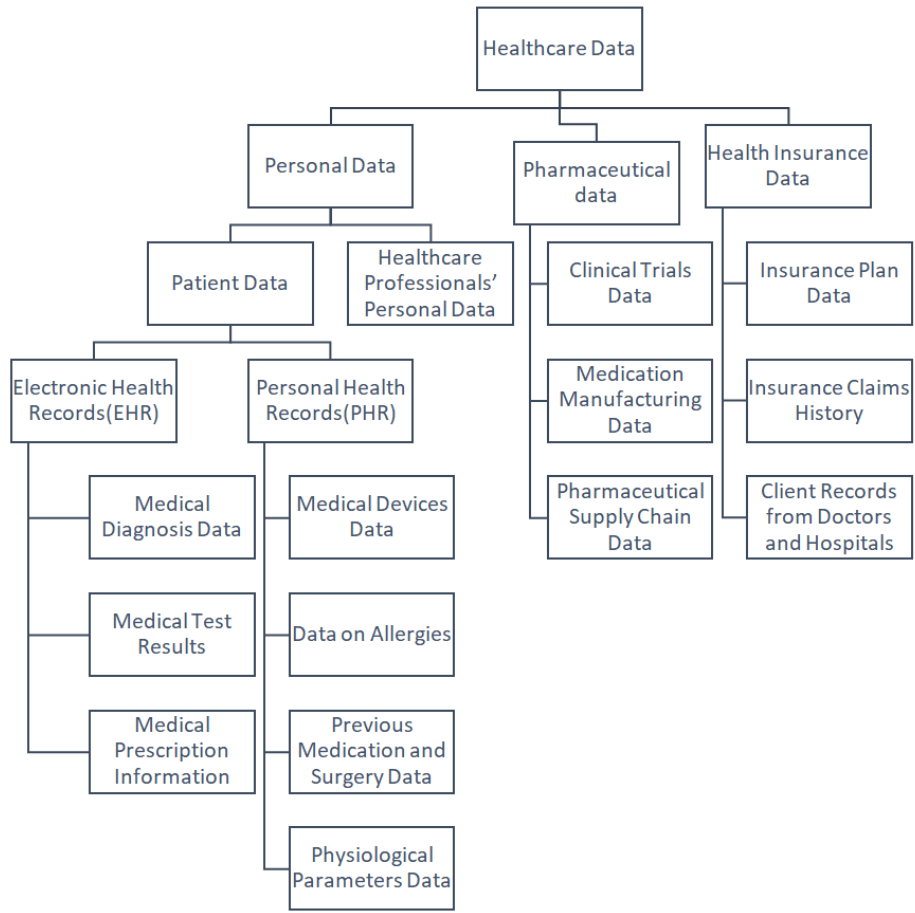

*Figure 1: Healthcare Data Types*

In summary, in this section, we discussed different healthcare data types (summarized in Figure 1: Healthcare Data Types) and the institutions that store and manage the same. Here, we see that the scope of healthcare data goes beyond just the medical diagnosis reports and personal health information and involves data belonging to pharma, health insurance, and details of healthcare professionals.

## 3 Healthcare data related services

In section 2, we detailed different healthcare data categories. Since the healthcare data goes beyond just the medical reports and the personal health information, the related services also go beyond diagnosis by doctors and medical tests. The scope of these services is diverse. They range from individual patients via diagnosis to organizational and national level via healthcare resource deployment. Apart from this, the functioning of the services related to healthcare research programs, pharmaceutical distribution, and health insurance procedures require diverse healthcare data as a vital input. The healthcare data-related services are broadly classified as medical-related services, distribution services, and commercial activity-related services. In this section, we discuss such healthcare data-related services in detail.

The most significant healthcare service (medical-related service) based on the healthcare data is the **patient diagnosis** and treatment plan. The doctors affiliated to a hospital or operating in an individual capacity assisted by nurses primarily provide such services (Hayrinen et al., 2008). The doctors thus require data from the sources such as EHRs and PHRs to prescribe treatment through methods such as medication and surgery.

**Healthcare Research and Development (R&D) plays a vital role in developing healthcare-related services (medical-related services)** based on healthcare data. Apart from the primary task of diagnosis, doctors and other researchers contribute to the healthcare research activities based on their domain expertise and daily experience (such as mobile health applications, as in (Brown et al., 2020)). For the pharmaceutical companies, research for new medications is an essential component of their operations, along with clinical trials for new medications. These trials generate large quantities of data vital to the discovery of new forms of medication and treatment. Data such as EHR and pharmaceutical research data in the healthcare domain are essential for improving the medication for various ailments, healthcare monitoring devices, and even the guidelines related to daily functioning lifestyle (Koh & Tan, 2005). Also, healthcare data is an essential input for artificial intelligence (AI) based research in the healthcare domain. For example, the use of surgical robots, which find their utility through minimally invasive and precise incisions during the surgery process (Bergeles & Yang, 2014). Another example is the application of IBM Watson, a supercomputer for clinical decision support systems (M. N. Ahmed et al., 2017). In addition to new treatments, healthcare data aids the research on discovering new diseases, bacteria, and viruses.

In addition to the data described in section 2, the aforementioned services also contribute a significant amount of healthcare data (Dash et al., 2019). In (Dash et al., 2019), the authors discuss the commercial **big data analytics services** platforms operating in the healthcare sector, such as Ayasdi[1], IBM Watson Health[2], and Enlitic,[3] that use and generate a significant amount of healthcare data for both medical-related and commercial services. These platforms provide services in a number of applications such as **healthcare monitoring, prediction, knowledge, and recommendation systems** (Bahri et al., 2019).

---

[1] Ayasdi Care is a software package for healthcare solutions with respect to providing efficient data-based patient care strategies for doctors and hospitals.
[2] IBM Watson Health combines human experts and AI to help healthcare researchers provide patient care decisions based on big data analytics.
[3] Enlitic company uses deep learning algorithms on clinical data to help radiologists with illness diagnosis.

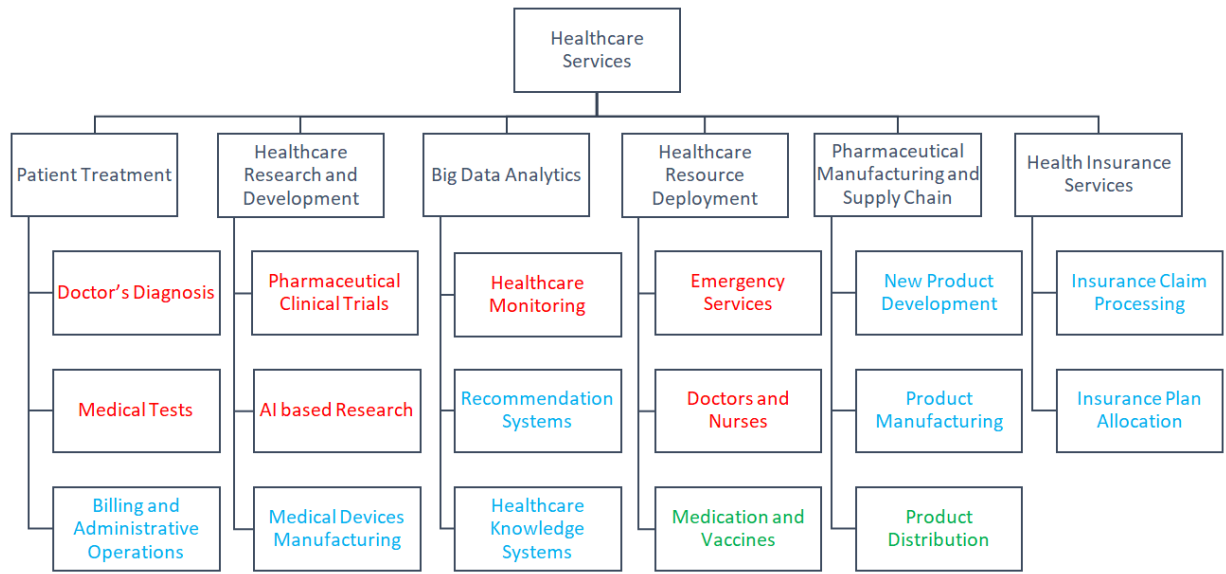

*Figure 2: Healthcare Data related Services. Here, color encoding represents the three classes: medical-related services (red), distribution services (green), and commercial activity related (blue).*

Apart from such services, there are services related to the **supply chain in the healthcare sector and resource deployment**. The EHR and PHR data of different patients from a specific geographical region consolidates to provide an overview of the health conditions prevailing in that region. Healthcare institutes and professionals deploy their resources to perform medical-related services based on the requirement deduced from the local EHR and PHR data (Landry et al., 2016). For example, the overwhelming number of patients due to community transfer of COVID-19 prompted an increase in intensive care units (ICU) capacity. The doctors and nursing staff were redeployed to provide medical services such as critical care units with immediate training (Lee et al., 2020). In another similar approach, in Ethiopia, the government deployed medical staff based on healthcare data such as child immunization, child mortality rate, maternal mortality rate, and presence of skilled manpower in different areas of the country (Teklehaimanot & Teklehaimanot, 2013). The resource deployment's objective was to achieve the results according to the World Health Organization's (WHO) Millennium development goals. The consolidated EHR and PHR data, along with pharmaceuticals manufacturing data, is an essential input for the pharmaceutical companies for the distribution services, marketing, manufacturing, sales, and distribution for healthcare-related products (Jaberidoost et al., 2013).

Another commercial activity service related to the healthcare sector is the **Health Insurance Service** (Pitacco, 2014). Different health insurance-related decisions such as premium amount, healthcare insurance plan, and whether to offer insurance or not for a patient is based on his/her healthcare data components such as medical history, ailments, allergies, and previous claim history.

In summary, in this section, we classified various healthcare data-related services into three classes (also summarized in Figure 2 using different color codes). We further identify that doctors play an essential role in diagnosis services. The non-medical professionals also form a significant part of the healthcare workforce and are involved in various other related services. All the aforementioned services require storage of different types of healthcare data as well as sharing amongst themselves. As such services are essential and critical, it is important that the healthcare-related organizations, which are both public and private, receive accurate data and take proper measures to secure the data they generate.

## 4 Security issues in conventional Healthcare data storage systems

In sections 2 and 3, we discussed different healthcare data types and the services built using them, respectively. Due to the sensitive nature of the data involved, its security is of utmost importance. As there is a significant increase in healthcare data generation, the data storage systems have transformed from paper-based record keeping to cloud-based storage. However, there are still security issues that are storage-centric, device-centric, and services-centric. In (Filkins, 2014), the author analyzed the cyber threat to healthcare organizations in terms of the level of compromise and malicious traffic. Most of such malicious activities mainly affect healthcare providers such as hospitals (72%), health insurance (6.9%), and pharmaceuticals (2.9%). The total cost of the compromised healthcare data, including the factors such as recovery, legal actions, and new security investments, was more than 142 million dollars in 2013 for the US healthcare sector. The compromise in PHR and health insurance data resulted in the loss of 12 billion dollars for 2 million US citizens in 2013. The main factors behind the compromise of healthcare data were identified as unsecured medical devices network, storage, and lack of data security awareness among general citizens. We discuss the security issues and challenges faced towards healthcare data storage and management as we proceed.

The storage-centric security issues arise primarily due to centralized storage and management by healthcare-related organizations and are further classified into vulnerabilities and threats. A significant security limitation with respect to centralized healthcare data storage systems is the **vulnerability due to a single point of control.** Since the healthcare data also consists of personal and critical information of a patient and healthcare professionals, such control points are always on the cyber-criminals target. Stealing such healthcare data and selling it illegitimately for financial gains is the prime motivation for cyber-criminals (Fugh-Berman & Ahari, 2007) (Hathaliya & Tanwar, 2020). A successful cyber-attack on the control point can compromise the data integrity and confidentiality of all the associated systems.

**Ambiguity in the patient data ownership causes privacy-**related issues and the problem of unauthorized access to stored data. For example, an EHR, although generated by the hospitals and the doctors, concerns a patient. Currently, such data is also managed by specialist third-party companies (Mandl & Kohane, 2012). Thereby it raises an issue of who owns the data (doctors, hospitals, or patients). Similarly, the patients' health insurance data is managed either by the insurance companies themselves or through specialist cloud-based storage companies. This brings up the privacy issue related to who owns the data (insurance companies or patients) and in what sense the data can be used. The ownership decisions related to healthcare data, such as access control, access rights, and modification privileges, are a significant issue affecting the security and privacy of healthcare data storage systems (Huda et al., 2009).

Due to such vulnerabilities, cloud-based healthcare data storage systems are under constant threat of various **cyber-attacks.** The attempt to gain unauthorized access is through different methods such as Denial-of-Service (DoS) attacks, password attacks, malware attacks, and social engineering attacks (Dogaru & Dumitrache, 2017). Attackers perform the Denial-of-Service attacks (Hussain et al., 2003) to shut down a hospital's network by flooding it with false requests. Similarly, attackers do a social engineering attack (Krombholz et al., 2015) to carry out insurance fraud and steal patient records through a careless insider. Further, attackers steal the passwords (Raza et al., 2012) through techniques such as brute force and phishing to gain unauthorized access to patient's healthcare records. Moreover, in the malware attack (Spence *et al.*, 2018), the attackers plant malicious code in either the healthcare data storage systems or medical devices to disrupt the functioning. A significant malware attack was the wannacry ransomware attack (Hsiao & Kao, 2018) which took place worldwide. It locked the files in the systems which were using the older Windows operating system and demanded payment of bitcoin cryptocurrency to unlock them. Since the beginning of the COVID-19 outbreak, the attackers have performed several similar attacks on hospitals (Muthuppalaniappan & Stevenson, 2021). In (Seh et al., 2020), the authors analyze financial losses suffered by the USA's healthcare sector due to data breaches owing to these attacks. They report that, in the last 15 years, around 249 million people were affected due to different data breach incidents. In 2019, 2013 incidents were reported in 86 countries. The average cost of a healthcare data breach is 6.45 million dollars worldwide, with cost in the USA being highest around 15 million dollars, according to an IBM report (Alder, 2019). Such a large-scale financial loss prompts proper measures to secure the healthcare system.

Apart from the attacks from the outside of the healthcare system, **malicious insiders** are also a significant issue threatening healthcare data security (Hathaliya & Tanwar, 2020). The attempt for unauthorized access to medical information is either carried out by someone inside the system or from outside the system taking advantage of a careless insider. An insider with the required credentials but not the authorization to see the healthcare data can misuse his/her privilege.

Apart from the storage-related issues, there are threats to healthcare data due to **vulnerabilities present in specialized or related healthcare devices** (such as CT scanners, X-ray machines, and MRI machines (Dogaru & Dumitrache, 2017) and those involved in mobile health or more commonly known as mHealth). A major device-centric security issue related to healthcare data protection is the vulnerability of such devices and the data generated using them. For example, the procedure to send the data involves connecting these devices to the cloud, where the aggregated information is stored and managed (Plachkinova et al., 2015). On top of, these devices are vulnerable to potential network-based attacks, such as false data injection attacks (M. Ahmed & Ullah, 2018), Denial of Service attack, and medjacking attack (Djenna & Eddine Saidouni, 2018). The individuals using these devices to record and send their physiological and healthcare data generally lack awareness about cybersecurity, making their device and data vulnerable. In (Kawamoto, 2017), the author analyses the incidents of data breaches in IoT devices through a survey where about 47.2% of the respondents from the healthcare sector accepted data breaches from their medical IoT devices.

In addition to the storage and device-related data security challenges, service-centric issues also cause healthcare data threats. Several issues arise in the big data analytics-based services in the healthcare domain (Abouelmehdi et al., 2018). These issues are due to parallel data storage and analysis across multiple servers. The computations on the same also run parallelly at multiple clusters. The compromise of even one such cluster can lead to wrong analysis (Gahi et al., 2016). Therefore, in addition to storage, the **vulnerability of computational analysis algorithms** is also a significant security issue.

In addition to medical data security issues, the protection of pharmaceutical supply chain data is also challenging. The pharmaceutical products are distributed to hospitals, clinics, and medical stores through a wide network of supply chains. For pharmaceutical companies, the protection of their supply chain information to prevent **pharmaceutical counterfeiting** is a significant challenge (Coustasse et al., 2010). The data on manufactured medications, including composition, manufacturing date, expiry date, and supply information, is vulnerable due to a single point of its storage and access. Pharmaceutical companies must mitigate the **risks of data breach attempts** from outside as well as inside the system to ensure the smooth operation of their services (Jaberidoost et al., 2013). Any breach of the product distribution information enables the activities such as **stealing for misuse** and counterfeiting of the drugs (Urciuoli et al., 2013). Therefore, the protection of pharmaceutical supply chain information is essential for the healthcare sector's proper functioning. The same applies to the companies producing biomedical devices for hospitals and doctors.

This section discussed healthcare data-related security issues around various healthcare affiliated organizations and their tasks (also summarized in Figure 3). We identify that the healthcare sector's vulnerabilities related to data security can either be storage-based, device-based, or services-based. The storage-based vulnerabilities arise primarily due to centralized control, while the device-based vulnerabilities owe their presence to the security resource-constrained specialized medical devices. The services-related vulnerabilities occur due to reasons such as parallel data storage and computation requirements. To mitigate the security threats caused by vulnerabilities, in the next section, we describe security parameters that are essential.

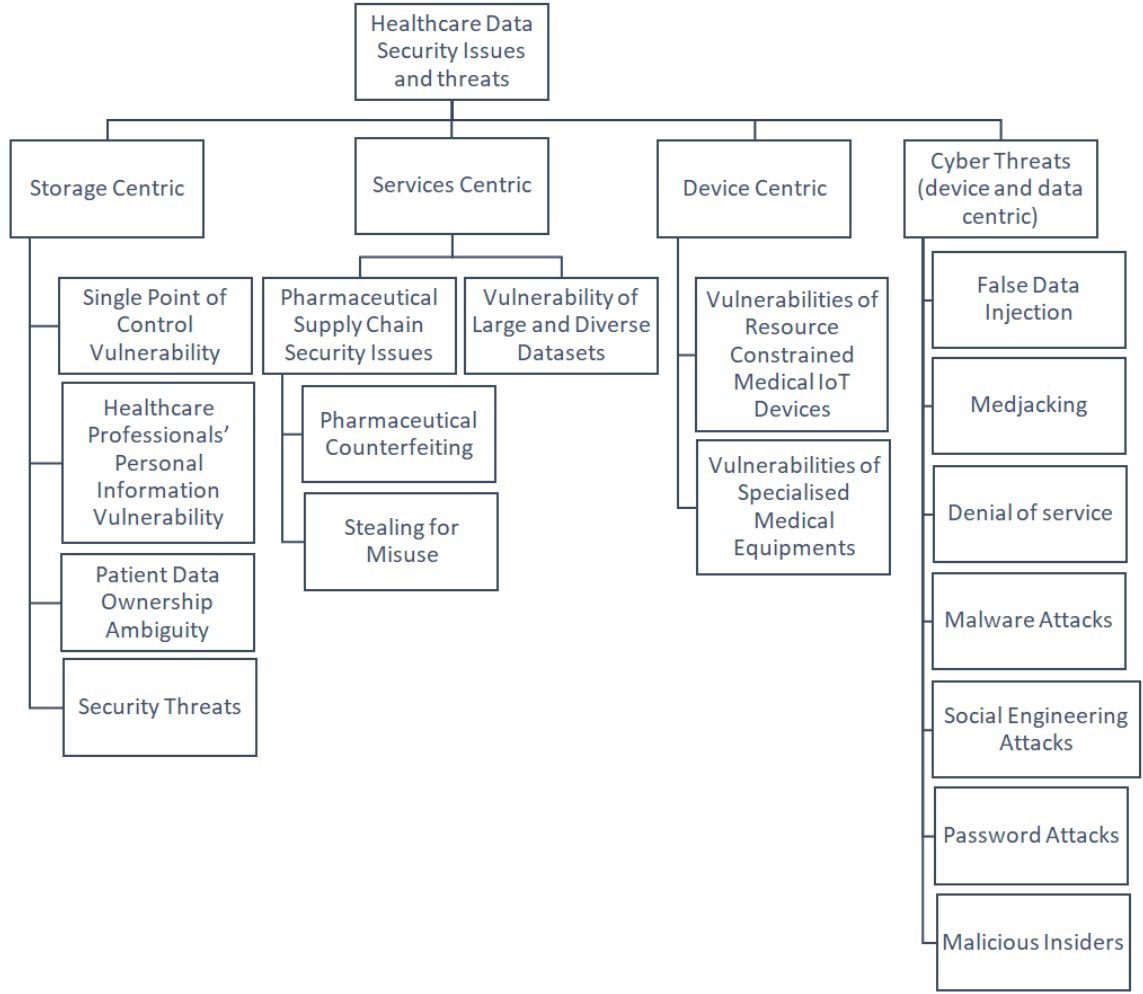

*Figure 3: Healthcare Data Security Issues*

## 5   Types of security parameters vis-à-vis healthcare sector

The security of healthcare data is essential for the stakeholders in the healthcare sector. The overall healthcare data protection encompasses a range of security parameters such as access control, data integrity, and traceability (Hardin & Kotz, 2019). In this section, we describe these parameters in detail.

An essential security parameter in the healthcare sector is **access control** to preserve healthcare data privacy (Abouelmehdi et al., 2017). The medical records of a particular patient should be accessed only by doctors who are involved in the treatment of the individual in question. Apart from the doctors, healthcare organizations' administrative staff access patient data from time to time for billing and insurance-related processing. Different countries have different data privacy laws, such as the HIPAA act in the USA specifically for healthcare data, general data protection laws like Personal Information Protection and electronic documents act in Canada, and General

Data Protection Regulation (GDPR) in European Union. These laws mandate the organizations storing healthcare data to get the concerned person's consent through a disclosure form before using his/her data and enforce the data privacy. They make access control of the healthcare data essential so that any unauthorized access can be prevented (Dagher et al., 2018). Therefore, the healthcare data storage and management systems should have proper access control measures to comply with the privacy laws enforced in the country in which they operate. The same applies to healthcare insurance companies, medical research organizations, pharmaceutical companies trying to access the medical data for their functioning and research activities. The consent of a patient for healthcare data access is also essential in such cases to prevent its misuse.

Currently, many patients consult different doctors across multiple hospitals either for different ailments or for taking a second opinion for the same condition. Further, various healthcare-related organizations collaborate for different data-related services. For example, the collaboration between the University of Oxford and AstraZeneca pharmaceutical company for developing a vaccine for prevention against COVID-19 (Mullard, 2020). Such aspects need a common platform where **data is secure, interoperable,** and has proper access control (Iroju et al., 2013).

In addition to access control, **preserving data integrity** is an essential parameter for healthcare data security (Pandey et al., 2020). Data integrity refers to the process of maintaining the accuracy and reliability of data over its entire utility cycle. In the healthcare domain, this includes maintaining the accuracy and integrity of the healthcare data, which, if lost, is catastrophic. Thus, the authorities should regulate any unauthorized modification in the healthcare data.

To identify the source of data in healthcare data storage and management systems, **traceability and auditability** (Hardin & Kotz, 2019) of the stored information are essential security parameters. Every day, a considerable amount of healthcare data either gets created or updated. Even after the proper measures, some of this data might be unauthorized and malicious. For instance, someone with credentials to access the healthcare data storage systems inserts incorrect information in the records file. Therefore, a secure healthcare data storage system must have the provisions for identifying the source of each new entry and modification in it.

Finally, **protection against cyber-attacks** is the most significant security parameter for healthcare data storage and management systems. There are several security threats to the healthcare sector in the form of cyber-attacks discussed in section 4. The objective of such attacks is diverse, including stealing data through unauthorized access, corrupting the data, making the network inaccessible, or causing system malfunctioning. Any successful attack causes a significant amount of damage due to the sensitive nature of the data. Therefore, the healthcare data management systems must be robust enough to thwart such attacks successfully.

In this section, we summarized the various security parameters that are essential while creating the healthcare data storage and management systems. The access control for healthcare data management systems should be strictly requirements-based such as doctors for diagnosis, administrative staff for billing to preserve healthcare data privacy. A proper design for inter-organizational operability should be there for the smooth operating of healthcare-related services without compromising on data security. Besides, healthcare organizations must have adequate auditing mechanisms for their stored data. Finally, the system must be robust enough to resist outsider as well as malicious insider attacks successfully. As we proceed, we discuss blockchain's utility in implementing these parameters for data security in the healthcare sector.

## 6 Blockchain components and properties for improvement in data security

Blockchain technology is based on decentralized networks introduced initially as a channel for cryptocurrency generation and payments (Nakamato, 2008). A blockchain is a distributed ledger consisting of data stored in the form of blocks sequentially linked to each other through hash values of block data. Once the data gets recorded on the ledger, it is hard to modify it due to it**s immutability property**. It **prevents unauthorized modification** as well as **provides for traceability** of healthcare data. There is no centralized storage of the blockchain data. Instead, each participating node in the decentralized network maintains a copy of the blockchain stored with it. At present, many applications use blockchain technology, including healthcare (Engelhardt, 2017). The use of blockchain in the healthcare sector has the potential to improve the transparency and security of healthcare data (Agbo et al., 2019).

The security of a blockchain network is linked to the permission level for new participants. Based on the permission rules for new participants joining, there are two categories of a blockchain-enabled platform: **permissionless** and **permissioned blockchain** (Wust & Gervais, 2018). In a **permissionless blockchain**, any person is allowed to join and transact with other users without any pre-authentication and a central authority. For **permissionless** blockchains, there is **no single point of control,** and hence the scenario of system malfunctioning due to the failure of a centralized administrator is not present. The blockchain for bitcoin cryptocurrency was the first instance of a permissionless blockchain (Nakamato, 2008). A permissionless blockchain is always public. On the other hand, a **permissioned blockchain** is for a single group or organization only. Here a **central authority is present** for the tasks such as providing permission to join, assigning roles, and granting privileges to the network's nodes. However, the other operations are carried out internally without interference from the authority. For example, Walmart Canada uses a permissioned blockchain for food supply chain management (Smith et al., 2020) using the Hyperledger Fabric framework (Androulaki et al., 2018). Since the number of nodes able to participate is limited in number, the time required to verify data and create a block is lesser than a permissionless blockchain. A permissioned blockchain is either a private permissioned blockchain or a public permissioned blockchain. A **private permissioned blockchain** has the criteria for entry as well as roles and specified tasks for each user. On the other hand, a **public permissioned blockchain** (Ruiz, 2020) has entry criteria but no discrimination over access. An example of a public-permissioned blockchain is Alastria (Ibáñez & Moccia, 2020), a non-profit multi-sector blockchain-based organization. A permissioned blockchain between multiple organizations is a **consortium blockchain** (A. Zhang & Lin, 2018).

The blockchain participants carry out the addition of new data in the form of blocks to the agreed-upon blockchain through the **consensus protocol**. The objective is to **get rid of a single controlling authority** with respect to decisions related to healthcare data storage and management. A proper consensus protocol should be resilient enough to function correctly against the issues such as failure of nodes, malicious nodes, delay in communication, and data corruption. In a consensus mechanism, a participant is selected either through a competition like solving a puzzle or through a randomized selection process. In (Xiao et al., 2020), the authors provide a detailed survey on consensus protocols in the blockchain. Among these protocols, **Proof of Work (PoW)** is the first significant consensus procedure (Nakamato, 2008). In this case, the participant nodes compete amongst themselves to find the solution to a cryptographic puzzle through a brute-force-based process. In another consensus protocol, **Proof of stake (PoS),** a node is selected from a group of validators for the block formation with selection probability depending on cryptocurrency units staked into the network (Saleh, 2021). Both PoW and PoS work in both permissionless and permissioned blockchains. **Practical byzantine fault tolerance (PBFT)** (Hao et al., 2018) and the **Raft consensus protocol** (Ongaro, D.; Ousterhout, 2014) are the consensus mechanisms designed only for the permissioned and consortium blockchain platforms with the property of byzantine fault tolerance (Castro & Liskov, 2002). Their functioning involves one node selected as the leader to initiate the block formation based on the feedback from the rest of the follower nodes regarding the validity of transactions. The objective is to reach a consensus with the correct information despite the malicious and faulty nodes present and sending false information. The blockchain-based networks use several other methods such as Proof of Authority (PoA), Proof of capacity (PoC), and Tendermint.

Another critical function of the blockchain is **hashing of the data through hash algorithms**. A hash algorithm converts a data string of any length into a fixed output of binary data. For blockchain, the hash algorithms used are SHA-256 (Gueron et al., 2011) and ethash (Han et al., 2019). It is nearly impossible to predict the input data from the hashed output. In a blockchain, each block has the previous block hash value stored as part of its data. The property of **hash-based linking of blocks** makes it difficult to alter the block data by predicting the ever-increasing chain of hash values. Apart from this, blockchain also uses public-key cryptography to create node addresses and digitally signing transactions. Some of the encryption techniques used for transactions in blockchain networks are elliptic curve cryptography (Hankerson et al., 2006), zero-knowledge proof (Partala et al., 2020), and ring signatures.

Blockchain technology is constantly evolving since its introduction. A significant stage in evolution is the conception of blockchain 2.0 through the Ethereum blockchain platform.

Blockchain 2.0 identifies and implements blockchain as a programmable decentralized trust platform (Kehrli, 2016) by introducing Smart Contracts (Ulieru, 2016). **Smart contracts** are self-enforceable autonomous programs containing predefined conditions for executing a transaction on Ethereum Virtual Machine. Smart contracts provide an agreement between the nodes participating in blockchain transactions. Once they get added to the blockchain, no one is able to modify them. When a smart contract is called for carrying out a transaction, each validating node executes it to verify the transaction's correctness and reach a consensus. Due to verification by each node, a smart contract should be deterministic and verifiable. The conditions executed through smart contracts get recorded as transactions during addition into the blockchain.

As discussed in section 4, there are different types of services such as medical diagnosis, conducting medical tests, and administrative tasks in healthcare. The assigning of different roles and privileges to the participating nodes based on which service they carry out is essential in this case. Therefore, permissioned blockchains are deemed more suited for healthcare data security. While implementing blockchain in the healthcare sector for data security, one requires several initialization operations. One such operation is assigning predefined roles and privileges to the entities such as doctors, staff, and patients. In addition to initialization, there are tasks related to the functioning after implementation, such as adding, modifying, and accessing healthcare data.

By defining healthcare data access control through smart contracts, blockchain **enables patient ownership and control** over his/her data. A patient is able to see the history of his data concerning the activities such as addition and access of the records. In addition to patients, if the different hospitals use blockchain for inter-organizational interaction and sharing of healthcare data, it **enables a secure standard format** for efficient processing of healthcare-related services. The integration of hospitals with pharmaceutical and health insurance companies through blockchain provides a secure platform for efficiently carrying out services such as medication distribution chain and insurance claim processing.

Estonia is one of the first countries that is officially incorporating blockchain in its healthcare sector functioning (Heston, 2017) to ensure data security and its ease of availability for patients. For doing so, Estonia collaborates with Guardtime (a private data security company) to secure its citizens' health records through blockchain. The government-authorized entities do the decentralized storage. Auditing is error-free due to the blockchain's immutability property. Within the blockchain, the medical researchers get anonymized data rewards in exchange for mining services.

Next, we answer the fundamental questions related to what, who, when, and why (4W) blockchain technology. We first answer "**what**" and "**why**" questions. For securing the healthcare data such as EHR, PHR, pharmaceutical data, health insurance data, and data generated by the services (discussed in section 3), blockchain technology is one solution.

Blockchains provide access control, integrity, traceability, and immutability to the healthcare data, which is currently missing from the traditional healthcare system that usually stores the data in a centralized manner. The healthcare institutions such as hospitals, clinics, pharmaceutical companies, and health insurance companies that manage data should implement blockchain-based solutions and are thus the ones **who** need blockchain to secure the healthcare data they are storing and the data their services are generating. Finally, we answer the **"when"** question. To carry out healthcare-related services, the people involved, such as doctors, staff members, and patients, perform different tasks such as uploading, modifying, and accessing healthcare data components. When used in the healthcare sector, blockchain securely performs these tasks, leading to improved healthcare data security.

In a blockchain, copies of data get distributed across the nodes of the network. This distribution, along with immutability, creates the problem of maintaining the privacy of data. The blockchain embedded encryption technique enables user privacy, but it does not guarantee the data's privacy. In the healthcare segment, as the data contains the medical history, diagnostic information, and medical test reports in the form of text and images, it has a large volume and is sensitive. For this reason, as a solution, only lightweight values such as hashes, pointers, and metadata are stored on the ledger. Additionally, **to enable privacy protection in a blockchain-based healthcare data storage system, different encryption and authentication techniques are used on top of blockchain-provided encryption.**

This section discussed the advantages of blockchain and its properties vital in establishing healthcare data security. The immutability of blockchain prevents any unauthorized modification of healthcare data. The ability to operate without a centralized authority by establishing a general majority-based consensus procedure provides for a transparent functioning of the healthcare sector. Proper use of smart contract functionality enforces the security parameters, such as access control for the healthcare-related services, inter-organizational operability of healthcare organizations, and traceability of healthcare data. In the next section, we answer **how** different state-of-the-art approaches implement blockchain for improving healthcare data security.

## 7  Blockchain for improvement in healthcare data security

In this section, we survey different proposed systems that implement blockchain to improve healthcare data security through its inherent properties and functionalities.

In (Zhuang et al., 2020), the authors propose a permissioned blockchain framework for creating a patient-centric health information exchange system with an objective to provide patients the **control** of their data. In the proposed system, the EHR reports are encrypted and stored on the respective hospital's servers, while their hash values are stored on the ledger to prevent tampering. The hospital administrators also create the pointers to each patient's EHR data referred to as touchpoints and store them on the ledger. The patient provides data access to a doctor by adding him/her to the "allowed list" through the smart contract. Any mismatch between the stored records and their hash values on the blockchain gets marked. Like (Zhuang et al., 2020), in (Shahnaz et al., 2019), the authors use the permissioned blockchain and smart contracts to **secure healthcare records by defining access rules**. The EHR data is stored using the interplanetary file system (IPFS) protocol (Benet, 2014), while the ledger stores the hash values of the EHRs. In the proposed system, smart contracts define patient, doctor, and administrator roles in the healthcare network and implement the health records' storage and management functions. A blockchain transaction in the proposed system defines the roles, adding, viewing, updating, and deleting health records. The administrator provides entry to the doctors in the system through a transaction. Once assigned as a doctor, a node performs the transactions related to adding, viewing, and updating the EHR data. The access control for the transaction activities is directly related to the defined roles. Any form of unauthorized tampering traces back to the user who performed the said transaction.

Similarly, in (Shen et al., 2020), the authors propose a system for the secure sharing of collaborative healthcare data. The system consists of three stakeholders, third-party, data owners, and miners operating on a consortium blockchain. Here, the third-party is referred to as the entities that perform data analysis-related tasks based on the requests and charge for them. The requested data is stored on the blockchain by the owners and accessed by the authorized third-party. The data owners and miners receive their share of services' generated revenue through smart contracts and a revenue model based on the Shapley value method (An et al., 2019). It provides for a secure and predefined incentive mechanism for healthcare data analysis services. In another similar work, in (Qiao et al., 2020), the authors propose a solution for the secure sharing of medical data between organizations from different regions using cross-chain communication between multiple consortium medical blockchains.

In addition to organization-centric data security, blockchain technology is also used for **secure storage and management solutions for home-based healthcare monitoring devices**. In (Li et al., 2020), the authors propose ChainSDI (Chain software-defined infrastructure), combining permissioned blockchain with SDI for healthcare to ensure regulation compliance by medical institutions. The compliance system is used for patient-generated sensor data from wearable devices in addition to EHR data. The authors claim that their system is compliant with the US government's Health Insurance Portability and Accountability Act (HIPAA). In the proposed system, the stakeholders are the admins that manage the data storage, data user (e.g., researcher), and patients. Their system invokes direct communication between a patient and the data user for access rights. The admin nodes verify whether the data user and patient are on the list registered on the blockchain. The admin node grants a data user's request to access data after receiving approval from the concerned patient. All these activities are enabled through smart contracts and get recorded on the ledger as transactions.

**Third-party independence** is another parameter using which healthcare data security is improved. In (S. Wang et al., 2019), the authors combine blockchain with attribute-based encryption and cloud storage to **enable PHR sharing without relying on a third-party**. The hash values of encrypted PHR data get stored on the blockchain, along with index linking to the patients' ID and data stored in the smart contract. The stakeholders of the system are patients, users, and cloud server organizations. The PHR data stored is encrypted using attribute-based encryption (Bethencourt et al., 2007). When a user requests a patient for data access, the patient generates attribute-based keys according to user requirements. This key enables the user to access the required data stored on the server. Therefore, it allows the patients to have requirement-based access control on their data without relying on a third-party.

Similarly, in (Madine et al., 2020), the authors propose a system for **patient-controlled secure PHR management** without dependence on the third-party. Their solution works on both permissioned and permissionless blockchains and uses an IPFS system for data storage. The patients, doctors, hospitals, and other healthcare-related organizations are registered on the blockchain as nodes through a controller smart contract. The request and permission for accessing records are enabled through smart contracts and get stored as transactions on the ledger. The system enables the patient to generate a symmetric key used to encrypt his/her PHR data. Then the blockchain-provided public key is used to encrypt the symmetric key. While granting access, the patient's private key is combined with the doctor's public key to generate a re-encryption key. This re-encryption key is used to encrypt the symmetric key. The symmetric key is then obtained using the doctor's blockchain-provided private key followed by the doctor accessing the requested data. Therefore, the proposed system provides an extra layer of data privacy and enhances security. The hash values of PHR are stored on the blockchain, and any unauthorized change gets detected. Similarly, in (Meena et al., 2019), the authors propose a hyperledger fabric-based healthcare ecosystem consisting of hospitals, insurance companies, pharmacies, and patients. The objective is to establish the patient's control over his/her data. A vital component of the proposed system is maintaining user and data privacy while forwarding the shared data in a computationally efficient manner through a proxy re-encryption algorithm.

Besides the techniques mentioned above, there are often emergencies when a patient is not in his/her senses to provide consent for access to PHR data. In such a case, access to the same is critical for emergency response staff and doctors involved in treatment, thereby needing **secure temporary access for emergency cases**. In (Rajput et al., 2019), the authors propose a Hyperledger fabric-based emergency access control management system (EACMS) for such situations. In EACMS, the patients grant emergency rights for access of their data to emergency doctors through smart contracts at the time of registration in the system. In EACMS, there are three registered stakeholders: a patient, a doctor, and an emergency doctor (ED), apart from the system administrator. Here a doctor is referred to as the regular doctor. During an emergency, the administrator assigns a registered ED to the concerned patient. After verifying the authenticity, the tasks done by ED, such as accessing patient records and doing modifications, are recorded as transactions on the blockchain and are immutable. There is no dependence on a third-party for emergency access procedures in the proposed system as the registered patients have already defined the access policy through the smart contract.

In (Y. Wang et al., 2019), the authors propose a consortium-based blockchain along with cloud-assisted storage of EHRs to prevent unauthorized access. Their proposed system has four stakeholders, the data owners (DO), which are patients; data providers (DP), which are doctors; cloud servers (CS); and data requesters (DR). The health records are encrypted and uploaded on the cloud by DP after getting authorization from the concerned DO. The DP creates the keywords and index of data and stores them on the blockchain. The DR requests data through the keywords provided by the DP to locate the relevant EHR index on the blockchain. Then DO authorizes the access request.

Besides access control and data management, blockchain also finds its usage for **securing mutual communication** between patients and data sharing between doctors from different hospitals (Liu et al., 2019). Here, a typical transaction stored on the blockchain consists of the medical data shared along with the granter and requester node's ID. In the approach, there are three stakeholders: the system manager, hospitals, and the users such as doctors and patients. This solution uses an improved version of delegated proof of stake (DPoS) consensus. In the consensus mechanism, the doctors act as delegates in DPoS while the hospital server acts as verifier supernode. For each doctor, a score based on shared records' authenticity is also calculated and stored on the ledger.

Apart from the access control, **secure inter-organization authentication** is an essential parameter for secure sharing of healthcare data because patients often visit multiple doctors and hospitals for either different ailments or to get a second opinion. In (Yazdinejad et al., 2020), the authors propose a public blockchain-based authentication system for distributed hospital networks. The hospital nodes act as validators and perform the task of adding members such as staff and patients to the blockchain and generating encryption keys for them. The creating and sharing of the keys are done based on the authentication of the concerned patient's device MAC address. The same encryption keys get used for migration across the hospitals. The transaction data recorded on the blockchain includes the patient health data linked with the registration information. Such a mechanism removes the dependency on the third-parties as well as decreases the time required for authentication.

Note that the solutions mentioned above do not consider the **direct participation of medical devices** in sharing the data. In (Zghaibeh et al., 2020), the authors propose SHealth, which overcomes this limitation and allows direct participation of medical devices. SHealth is a private multi-layered Hyperledger based blockchain system. Here, the nodes are classified into five groups based on the privileges. Group 1 includes partners as the government, while group 2 consists of providers such as hospitals and pharmacies. Group 3 comprises all individual users such as doctors, nurses, and patients. While nodes that belong to group 4 and group 5 nodes are the addresses of stored records and IoT terminals, respectively. User privileges assigning and full data storage are done by nodes in group 1, while group 2 nodes maintain the blockchain and participate in the consensus process. The group 3 nodes are the service consumer nodes and use smart contracts related to health records access. All the stakeholders access the relevant data without compromise on authenticity. Patients also use smart contracts for requests and queries related to appointments, prescriptions, and medical history. The proposed system enables the confidentiality of the activities such as doctor visitation and enables secure and smooth data portability among healthcare organizations.

In (Nguyen et al., 2019), the authors discuss the security of medical data exchanging through mobile devices by enabling blockchain-based transactions. In the proposed system, the mobile data (i.e., EHR, PHR, wearable devices data) gets stored in a distributed IPFS file system. The patient nodes are provided with an area ID based on their location, while each area has an EHR manager node. The blockchain stores the tuple (containing patient ID, patient area ID, transaction type, and mobile data hash). The transaction types involve uploading, modification, and access of EHR data. Smart contracts define and enable the functioning of these transactions. In the proposed system, admin nodes provide access to the EHR manager nodes and define their access roles. EHR managers upload encrypted EHR data of their patients in the IPFS file system after receiving data. The data storing process gets recorded as the transactions.

Other than the access control, **the interaction between wearable devices and associated smartphone applications also needs to be secured**. In (Tomaz et al., 2020), the authors provide a lightweight non-interactive zero-knowledge proof authentication mechanism for resource-constrained mobile healthcare devices. The proposed blockchain-based system shares medical data for remote health monitoring along with the provision of attribute-based encryption for patient-controlled data privacy. The metadata of health records is stored on the blockchain, while the authenticity of accessing nodes is established through smart contracts.

In (J. Xu et al., 2019), the authors propose a blockchain-based solution called Healthchain for **protecting the privacy of large-scale health data** uploaded through medical IoT devices. It consists of two sub-blockchains, Userchain, which is public, and Docchain, which is a consortium blockchain. The Docchain records doctors' diagnoses and contains the transactions regarding the same. The userchain records the transactions related to uploading medical IoT devices data and accessing own diagnosis records. The transactions on these chains include data from IoT devices, keys for access control, and doctors' diagnoses. It protects the interest of patients by keeping the information confidential and ensuring accountability through record keeping.

The healthcare sector is regarded as an essential service. The healthcare data is sensitive and crucial for the smooth functioning of medical facilities. Such aspects cause the healthcare data storage and management systems to be constantly under attack from outside as well as inside the system. The objective of such attacks varies from data theft to denial-of-service. The threat is much more severe in the case of medical IoT device networks. In (Meng et al., 2020), the authors propose a blockchain-based solution for **protection against malicious attacks**. Here, they focus on collaborative Intrusion Detection Systems (IDS) based trust management for preventing insider attacks. Each user has the IDS system installed on its device, having traffic monitor components, communication components, and a blacklist of malicious nodes. In the proposed system, there are two layers: the medical smartphone network (MSN) layer and the chain layer. The MSN layer deals with the interaction of smartphone devices with centralized storage. The chain layer consists of a consortium blockchain that helps users upload the unwanted and malicious feature/data packets information. Based on the provided information, the central server calculates dynamic trust values through Bayesian inference (Sun et al., 2006). Based on the trust values, each user can record its own blacklist of malicious nodes. The data recorded on the blockchain is immutable, and hence the list of malicious nodes cannot be manipulated by anyone.

In (Zhu et al., 2020), the authors propose a solution to **prevent counterfeiting** medication information in the supply chain. Here, blockchain technology is used to track the medication distribution within the nodes comprising manufacturers, distributors, pharmacies, hospitals, doctors, and regulators. If an issue gets detected at any stage of the supply chain, it is traced to the source of origin using the blockchain's stored information.

In another solution targeting malicious activities (Saldamli et al., 2020), the authors use blockchain for **insurance fraud detection**. The authors point out the lack of coordination between health insurance companies with respect to data sharing and show that most of the financial loss is due to instances of fraud claims through fake healthcare information. In (Saldamli et al., 2020), the authors propose linking insurance claims data using BigchainDB to create a distributed database on blockchain concepts. For the patients, they propose two different identifications, patient ID and billing ID. The billing ID is linked to a patient ID and is unique for each healthcare service the patient has availed. Whenever a health insurance company processes a claim with respect to a billing ID, it gets recorded on the ledger. Then the amount in the claims file and billing file is compared to check for authenticity of the insurance claim. Due to blockchain data's immutability, it is impossible to avail insurance for the same billing ID twice. However, the proposed work assumes that the billing information generated from the healthcare service provider is authentic. Therefore, the development of methodology of verifying medical bills' authenticity is a potential future research problem.

We provide a summary of the blockchain-based solutions for healthcare data security in **Table 1**. In **Table 1**, we summarize and discuss the specific issues present in the state-of-the-art solutions. Besides the specific issues, there are general blockchain-based limitations that are common across all the proposed solutions. We discuss those limitations in the next section. In summary, in this section, we discuss how blockchain technology is applied to achieve healthcare data security. The use of blockchain enables security-based functionalities such as access control, inter-organizational operability, and preventing counterfeiting. Regarding user and data privacy, proposed blockchain-based systems use additional data encryption techniques for enabling privacy preservation.

| Approaches | Blockchain | | Functionality | | | | | | | | Issues |
|---|---|---|---|---|---|---|---|---|---|---|---|
| | Types | Storage | Storage | Exchange | Independence | Incentives | Emergency | Privacy | IoT Device-based | Malicious detection | |
| (Zhuang et al., 2020) | $P_r$ | ⊥ | ✓ | ✓ | ✓ | | | | | | Setup is required to be installed at each healthcare facility. |
| (Shahnaz et al., 2019) | - | ⊥ | ✓ | ✓ | ✓ | | | | | | No defined specification of organizational involvement. |
| (Nguyen et al., 2019) | $P_r$ | ⊥ | ✓ | ✓ | ✓ | | | | ✓ | | Solution for low-network latency in wearable devices networks not provided. |
| (Shen et al., 2020) | C | † | ✓ | ✓ | ✓ | ✓ | | | | | Multiple transactions are required for each sharing request due to the massive volume of data. |
| (Qiao et al., 2020) | C.C | ⊥ | ✓ | ✓ | ✓ | | | | | | The analysis provided is on a limited scale with few number of nodes. There is no clarity on real-time performance and the cost of building the system. |
| (Li et al., 2020) | C | ⊥ | ✓ | ✓ | ✓ | | | | ✓ | | Databases storing healthcare data are assumed to be privacy laws (HIPAA) compliant. |
| (S. Wang et al., 2019) | - | ⊥ | ✓ | ✓ | ✓ | | | | | | There is no mechanism to verify cloud server response to the requirement related to PHR data management. |
| (Madine et al., 2020) | $P_u+P_r$ | ⊥ | ✓ | ✓ | ✓ | | | | | | Multiple deployments of the same smart contract are required for it to work across different countries. No integration of a patient's PHR in case of consultations |

| | | | | | | | | | | |
|---|---|---|---|---|---|---|---|---|---|---|
| | | | | | | | | | | in other countries. |
| (Rajput et al., 2019) | $P_r$ | ↓ | ✓ | ✓ | ✓ | | ✓ | | | The system assumes that the person given access to PHR data in case of emergency is not malicious. |
| (Y. Wang et al., 2019) | C | ↓ | ✓ | ✓ | ✓ | | | | | The computational performance of smart contract algorithms depends on the length of EHR data keywords (pointers) stored on the blockchain. |
| (Zghaibeh et al., 2020) | $P_r$ | ↓ | ✓ | ✓ | ✓ | | | ✓ | | A new node joins through a bureaucratic procedure that involves government nodes physically verifying the candidate against a government database. A patient can get his/her health record updated through a partner, which refers to a health insurance organization. |
| (Liu et al., 2019) | $P_r$ | ‡ | ✓ | ✓ | ✓ | | ✓ | | | Medical data to be shared must be lightweight due to on-chain sharing. |
| (Tomaz et al., 2020) | - | ↓ | ✓ | ✓ | ✓ | | ✓ | ✓ | | The proposed system works on the assumption that the initial device registration is done in a secure environment. |
| (J. Xu et al., 2019) | $P_u$+C | ↓ | ✓ | ✓ | ✓ | | ✓ | ✓ | | The system assumes that any adversary working to crack the encryption key has limited computing power. Also, the off-chain channel between patients and medical IoT devices is assumed to be secure. |
| (Meena et al., 2019) | $P_r$ | ‡ | ✓ | ✓ | ✓ | | ✓ | | | The money transfer is done outside the network, while the insurance and pharmacy- |

| | | | | | | | | | | | |
|---|---|---|---|---|---|---|---|---|---|---|---|
| | | | | | | | | | | | related bills are stored on the ledger. |
| (Yazdinejad et al., 2020) | P$_u$ | ‡ | ✓ | ✓ | ✓ | | | | | | Patient control over his/her healthcare data is not present. |
| (Meng et al., 2020) | C | ⊥ | ✓ | ✓ | ✓ | | | | | ✓ | The impact of external attacks such as DoS on the system is not considered. Also, the experiment and analysis are done with a very limited number of nodes. |
| (Zhu et al., 2020) | P$_u$+P$_r$ | ‡ | ✓ | ✓ | ✓ | | | | | ✓ | The proposed system assumes that the medication information uploaded on the blockchain is authentic, and each node is honest. |
| (Saldamli et al., 2020) | P$_u$ | ⊥ | ✓ | ✓ | ✓ | | | | | ✓ | The blockchain technology used (BigchainDB) does not have a stable version and community support. |

*Table 1: List of Proposed state-of-the-art blockchain-based solutions for healthcare data security. Here, P$_r$: Private blockchain, P$_u$: Public blockchain, C.: Consortium based blockchain, C.C.: Cross Consortium based blockchain, -: "no mention of", ⊥: only hashes and transactions stored on-chain, †: only shared data and transactions stored on-chain, and "‡": all data and transactions stored on-chain.*

## 8 Technical limitations in the implementation of blockchain in healthcare

For healthcare data security, blockchain's use provides a significant edge over cloud-based data storage and management systems. However, its implementation in the healthcare sector has several challenges and limitations. Some of these challenges are generic, i.e., applicable to blockchain regardless of the application, while some are healthcare-sector specific. This section discusses the technological limitations in the current blockchain-based systems concerning the healthcare sector.

The **scalability issues for the large networks** are one of the blockchain technology's functional limitations (Yli-Huumo et al., 2016). This issue is much more severe in the case of healthcare data storage. In the healthcare sector, the data generated is much more in volume than financial transaction data due to the presence of images such as X-Ray and CT scans (Esposito et al., 2018). Storage of such a large amount of data at every node is cumbersome. To resolve such an issue, in (Xia et al., 2017), the authors propose to store only partial data such as metadata, hash values, and pointers on the blockchain while other data on the servers.

Apart from scalability, another issue is healthcare **data privacy** (Nawari & Ravindran, 2019). As a blockchain operates in a distributed network, each node stores a copy of the ledger. However, it is not in the interest of a patient to have copies of his/her medical diagnosis reports shared across the network. Therefore, the issue of data privacy is also one of the reasons for a blockchain-based hybrid data storage system in the healthcare sector. However, even if whole data is not shared across the network, the transaction information linked to the nodes' ID is available for everybody on the blockchain. Therefore, the blockchain by itself cannot protect the privacy of activities carried out by nodes. As we discussed in the previous section, several encryption techniques are applied in addition to blockchain-based encryption to achieve user privacy. Various policies related to access control are defined in state-of-the-art blockchain-based systems to address data privacy limitations. Although these policies get embedded into the system through storage on the blocks, there is **no way to enforce them within the network** without functionalities outside the blockchain properties (Hardin & Kotz, 2019). To handle such limitations, some nodes are given more privileges than the rest by giving them administrative powers. Such a requirement of providing some nodes more privilege than the others is one reason why most of the proposed frameworks for healthcare data security are based on permissioned blockchains. This leads to **centralization** within the blockchain-based networks (Nawari & Ravindran, 2019).

Apart from the overall functionality limitations, there are also specific **limitations associated with various consensus algorithms** used in blockchain (S. Zhang & Lee, 2020), which are generic and equally applicable to blockchain in healthcare applications. The proof of work needs computational resources, which in most cases is outside the scope of individual patients as well as some small hospitals. The requirement of computational resources works against the principle of participants' equality, even for the permissioned blockchains. For the PBFT consensus used in Hyperledger blockchain, the large number of messages required to be sent between the nodes causes network congestion. Nowadays, patients send their personal healthcare data recorded through various medical IoT devices and smartphones. The requirements for full participation force these resource-constrained devices to be mere spectators in the blockchain networks. Similarly, proof of stake consensus favors financially strong healthcare institutions. The disadvantage of such favor gives disproportionate power to big institutions with respect to discretion regarding inclusion or exclusion of certain transactions.

The functions like access control, privacy, recording, modification, and seeing the healthcare data get executed through smart contracts. Smart contracts play an essential role in the automation of tasks. However, there are several **smart contract limitations** (Zou et al., 2019) that directly affect the functioning of healthcare data security**.** Once a smart contract gets recorded on the blockchain, no one can change its code. Due to this, the developers need to check for vulnerabilities before deploying. Anytime these contracts are called to execute a task, every validating node runs it to verify the transactions. It again creates a privacy issue as every node has access to all the data used by the code. Therefore, extra care must be taken while developing a smart contract in deciding how much data and the encryption keys to provide.

Due to the above limitations, there are several **security risks** associated with blockchains, such as double spending (a participant creates a parallel transfer of the same data to two different participants leading to an invalid transaction), 51% attacks (occur when a participant or a group controls most computational resources and tries to tamper with the blockchain data), and risks associated with encryption techniques used in blockchain (Nawari & Ravindran, 2019). The risks related to the encryption techniques are due to the progress of computational technology. For instance, quantum computing's fast progress has the potential to make current blockchain encryption techniques obsolete against quantum attacks (Fernandez-Carames & Fraga-Lamas, 2020).

In this section, we discussed the technology-related limitations of blockchain implementation for healthcare data security. These include the blockchain's limitations in general and the constraints due to the healthcare data structure and functioning. The two main issues are scalability and data privacy, for which additional measures such as hybrid storage model and re-encryption of data are additionally applied, as discussed in the previous section.

## 9 Regulatory challenges for blockchain in healthcare data security

Apart from the technology-based limitations, there are several legal and regulatory challenges while applying blockchain to improve healthcare data security. From the perspective of governments and their relevant departments, there are many issues to be resolved.

A significant one amongst these is related to the issue of privacy. If an individual patient or an organization decides to leave the blockchain-based network, their previous data can not get erased due to its immutability property. It goes against the structure of the "**right to be forgotten**" granted under privacy rights in most countries (Gabison, 2016). This right explicitly enables an individual to get his/her data removed in the event of opting out of the organization. It creates a regulatory challenge for the stakeholders involved.

Blockchain-based healthcare data storage systems are **not yet officially standardized** (Yeoh, 2017). The different state-of-the-art systems get adopted by the organizations following their requirements. The healthcare organizations using blockchain have their specific data storage format, encryption techniques, and consensus algorithms. It leads to **interoperability issues** across the blockchains, causing it to become difficult for hospitals and other healthcare organizations to collaborate. It becomes an issue for the patients due to the need to migrate their data across the chains. Therefore, standard guidelines for blockchain operations are warranted.

There is also the **issue of jurisdiction for permissionless blockchains** that go beyond national boundaries (Yeoh, 2017). It is a general issue of blockchain but is equally affecting healthcare as any other sector. In the systems operating through permissionless blockchains, a patient can get a diagnosis from a doctor of different countries, and the information gets stored on the ledger. The jurisdiction regarding ownership of healthcare data, payment information, and taxation is not defined for such cases. Although only the metadata or hash values get stored on the blockchain in most cases, the issue of ownership is still present. To avoid this issue, most of the proposed systems run on permissioned blockchains. However, even for permissioned blockchains, a clear regulatory code specifying the **laws of functioning, data ownership, compliance, etc., is warranted** for most nations. It reduces blockchain to an experimental and assisting technology for existing healthcare data management systems.

A significant issue in the way of widespread adaptation of blockchain for healthcare data security is the **lack of training and awareness** (Kramer, 2019). From the perspective of healthcare organizations such as hospitals, they have their share of concerns regarding the adoption of blockchain for data security. For migrating to the blockchain, the patients, doctors, and administrative staff need to undergo training to learn the secure usage for data storage and access. With the technology still evolving, the proper user interfaces for blockchain-based systems are still not in place, making it challenging for non-technical personnel to adjust to its functioning. The stakeholders must learn about the security aspects like the use of encryption keys, calling proper smart contract functions, and protection of their information to ensure data security. Hence, adequate training is essential for the widespread adoption of blockchain.

Overall, the healthcare organizations, government departments, and blockchain developing organizations need to coordinate amongst themselves for making blockchain a mainstream technology in healthcare.

## 10 Blockchain role in pandemics like COVID-19 for large-scale data security and management

In current times, all the nations are struggling to cope with the pandemic caused by coronavirus disease (COVID-19), known as severe acute respiratory syndrome coronavirus-2 (SARS-CoV-2). It first got detected in the Wuhan region of China in December 2019. Since then, it has found its way into almost all nations worldwide due to its infectious nature of spreading through close contact and respiration. It was declared a worldwide pandemic by the WHO in March 2020 (Kalla et al., 2020). At the time of writing, the total detected infections had crossed 100 million worldwide, causing more than 2 million deaths. It has caused an unprecedented strain on the economies worldwide, with the healthcare sector being most affected. To counter and wipe out the virus, efforts are being made on all fronts in healthcare, including emergency services, treatment facilities, medication, and research on dealing with such pandemics through vaccination. All the mentioned fields require dealing with the data for their functioning. The data includes the components, such as patient's health records, data on hospital resources, information on persons infected with coronavirus, research data for developing medication and vaccines. The security of this data generated on such a large-scale is a challenge for governments, healthcare organizations, and research laboratories. In this regard, blockchain performs a key role in managing the COVID-19-related healthcare data.

A significant application where blockchain finds its use is contact tracing for COVID-19. A patient's anonymity is preserved using blockchain-generated IDs while alerting the persons who were in close contact through Bluetooth-based technology. For this purpose, the authors in (H. Xu et al., 2021) propose a blockchain-based contact tracing solution named BeepTrace.

The traceability property also provides for tracking passenger movements. To facilitate the same, a blockchain-based solution for issuing immunity certificates has been proposed (Hasan et al., 2020). Using smart contracts, it issues medical passports for people resulting in a negative in the COVID-19 medical test. Combined with an interplanetary file system, it ensures the security of patient data. Since blockchain data is immutable, it cannot be manipulated by either of the stakeholders. Apart from this, by using smart contracts, the procedure for health insurance approval is made faster with lower processing costs.

A blockchain-based app Civitas (Wright, 2020), is used in Canada to control the impact of COVID-19. It maps people to blockchain to find out whether they are in quarantine or not. It also aids with the ideal time to go out of the house for essential tasks.

Besides, blockchain also aids in research for finding medication and vaccines to prevent and treat coronavirus disease. Currently, the medical records for COVID-19 are managed by each healthcare institution independently. In (Yu et al., 2021), the authors provide a blockchain-based secure data sharing platform for carrying out collaborative medical research and clinical trials. The hospitals and medical research institutions are the nodes in the decentralized network. All data is attached to blockchain-based pseudonyms instead of real-world identities to protect the patient's privacy.

Overall, blockchain plays a vital role in ensuring the security of healthcare-data-related operations for pandemic situations. It prevents information manipulation and provides the patients with the control of their data without relying on an independent third-party.

## 11 Conclusion and Perspective on Future Research Directions

The decentralization of systemic functioning has become a major objective for healthcare systems to achieve complete security, privacy, and accountability for the relevant data. The use of blockchain-based peer-to-peer networks plays a significant role in accomplishing the same. The requirement of cryptographic verification and majority consensus before adding new blockchain blocks brings transparency and joint accountability in the healthcare system where sensitive patient data is involved. It has several benefits to the patients in terms of monitoring the access and use of their data. It also streamlines the operations of doctors, healthcare institutions, and medical research centers in terms of acquiring the relevant information by removing the bottlenecks associated with the system's centralized functioning. For the security of the data, blockchain puts the responsibility on all the stakeholders through individual encryption and removes the need for a trusted third-party.

However, none of the proposed blockchain-based networks functioning in the healthcare sector are entirely decentralized. These systems have interference in the form of administrative nodes, which warrants the research required to achieve pure decentralization for full transparency. Blockchain-based systems in the healthcare sector also need a sustainable incentive generation scheme and sharing for the miners/validators to keep the network running. Specific to the healthcare sector, it is currently cumbersome to store all the data on the blockchain. Besides, data privacy is equally crucial. Therefore, instead of storing all healthcare data, only the metadata and hash values are stored on the blockchain. For data privacy in such conditions, additional data encryption gets applied above the blockchain-provided encryption. With the requirement of continuous research on scalability and encryption techniques, blockchain-based networks can play a vital role in data security for next-generation healthcare systems. Apart from the technology-based challenges, there are regulatory questions such as jurisdiction regarding ownership of healthcare data and a standard inter-organizational functioning format requirement.

Technologically, we need an overhaul in the blockchain structure and functioning, specific to the healthcare sector's data types, functioning hierarchy, and security issues. Our recommendation for the researchers and developers working towards achieving healthcare data security through blockchain is to first collaborate with the policymakers, healthcare organizations, and different stakeholders on a large-scale (e.g., country-scale) to understand the ever-changing global needs rather than providing healthcare data security solutions for a specific organization with a particular need. As we understand, data privacy is a significant issue to be addressed to achieve healthcare data security. It warrants the need to incorporate additional privacy measures within blockchain architecture. One research direction that researchers could focus on is to modify the blockchain architecture to include a privacy module. Besides, blockchain functioning is computationally expensive. Developing countries or third-world countries lack the resources for such a large-scale implementation of blockchain. Another research direction is to develop a resource-efficient blockchain that could guarantee the privacy of healthcare data as well. Also, a properly defined incentive scheme is required for the miners/validators to motivate them towards active participation. These incentives could be altruistic or financial aid from the government. The researchers could look into different incentive schemes and revenue sharing models. Finally, the acceptance of blockchain for healthcare data security depends on the people's willingness to adopt it as a solution. Therefore, proper training and awareness among the stakeholders are required for it to become a success.

## Acknowledgements

This work is partially funded by the National Blockchain Project at IIT Kanpur sponsored by the National Cyber Security Coordinator's office of the Government of India and partially by the C3i Center funding from the Science and Engineering Research Board of the Government of India.

## *References*

Abouelmehdi, K., Beni-Hessane, A., & Khaloufi, H. (2018). Big healthcare data: preserving security and privacy. *Journal of Big Data*, *5*(1), 1–18. https://doi.org/10.1186/s40537-017-0110-7

Abouelmehdi, K., Beni-Hssane, A., Khaloufi, H., & Saadi, M. (2017). Big data security and privacy in healthcare: A Review. *Procedia Computer Science*, *113*, 73–80. https://doi.org/10.1016/j.procs.2017.08.292

Agbo, C., Mahmoud, Q., & Eklund, J. (2019). Blockchain Technology in Healthcare: A Systematic Review. *Healthcare*, *7*(2), 56 (1-30). https://doi.org/10.3390/healthcare7020056

Ahmed, M. N., Toor, A. S., O'Neil, K., & Friedland, D. (2017). Cognitive Computing and the Future of Health Care Cognitive Computing and the Future of Healthcare: The Cognitive Power of IBM Watson Has the Potential to Transform Global Personalized Medicine. *IEEE Pulse*, *8*(3), 4–9. https://doi.org/10.1109/MPUL.2017.2678098

Ahmed, M., & Ullah, A. S. S. M. B. (2018). *False Data Injection Attacks in Healthcare* (pp. 192–202). https://doi.org/10.1007/978-981-13-0292-3_12

Alder, S. (2019). *2019 Cost of A Data Breach Study Reveals Increase in U.S. Healthcare Data Breach Costs*. HIPAA JOURNAL. https://www.hipaajournal.com/2019-cost-of-a-data-breach-study-healthcare-data-breach-costs/

An, Q., Wen, Y., Ding, T., & Li, Y. (2019). Resource sharing and payoff allocation in a three-stage system: Integrating network DEA with the Shapley value method. *Omega*, *85*, 16–25. https://doi.org/10.1016/j.omega.2018.05.008

Androulaki, E., Barger, A., Bortnikov, V., Cachin, C., Christidis, K., De Caro, A., Enyeart, D., Ferris, C., Laventman, G., Manevich, Y., Muralidharan, S., Murthy, C., Nguyen, B., Sethi, M., Singh, G., Smith, K., Sorniotti, A., Stathakopoulou, C., Vukolić, M., … Yellick, J. (2018). Hyperledger fabric: a distributed operating system for permissioned blockchains. *Proceedings of the Thirteenth EuroSys Conference*, 1–15. https://doi.org/10.1145/3190508.3190538

Bahri, S., Zoghlami, N., Abed, M., & Tavares, J. M. R. S. (2019). BIG DATA for Healthcare: A Survey. *IEEE Access*, *7*, 7397–7408. https://doi.org/10.1109/ACCESS.2018.2889180

Beigel, J. H., Tomashek, K. M., Dodd, L. E., Mehta, A. K., Zingman, B. S., Kalil, A. C., Hohmann, E., Chu, H. Y., Luetkemeyer, A., Kline, S., Lopez de Castilla, D., Finberg, R. W., Dierberg, K., Tapson, V., Hsieh, L., Patterson, T. F., Paredes, R., Sweeney, D. A., Short, W. R., … Lane, H. C. (2020). Remdesivir for the Treatment of Covid-19 — Final Report. *New England Journal of Medicine*, *383*(19), 1813–1826. https://doi.org/10.1056/NEJMoa2007764

Benet, J. (2014). *IPFS - Content Addressed, Versioned, P2P File System*. 1–11. http://arxiv.org/abs/1407.3561

Berenson, A. B., & Rahman, M. (2011). Prevalence and Correlates of Prescription Drug Misuse

Among Young, Low-Income Women Receiving Public Healthcare. *Journal of Addictive Diseases*, *30*(3), 203–215. https://doi.org/10.1080/10550887.2011.581984

Bergeles, C., & Yang, G.-Z. (2014). From Passive Tool Holders to Microsurgeons: Safer, Smaller, Smarter Surgical Robots. *IEEE Transactions on Biomedical Engineering*, *61*(5), 1565–1576. https://doi.org/10.1109/TBME.2013.2293815

Bethencourt, J., Sahai, A., & Waters, B. (2007). Ciphertext-Policy Attribute-Based Encryption. *2007 IEEE Symposium on Security and Privacy (SP '07)*, 321–334. https://doi.org/10.1109/SP.2007.11

Bietz, M. J., Bloss, C. S., Calvert, S., Godino, J. G., Gregory, J., Claffey, M. P., Sheehan, J., & Patrick, K. (2016). Opportunities and challenges in the use of personal health data for health research. *Journal of the American Medical Informatics Association*, *23*(e1), e42–e48. https://doi.org/10.1093/jamia/ocv118

Brown, C., Chauhan, J., Grammenos, A., Han, J., Hasthanasombat, A., Spathis, D., Xia, T., Cicuta, P., & Mascolo, C. (2020). Exploring Automatic Diagnosis of COVID-19 from Crowdsourced Respiratory Sound Data. *Proceedings of the 26th ACM SIGKDD International Conference on Knowledge Discovery & Data Mining*, 3474–3484. https://doi.org/10.1145/3394486.3412865

Burbidge, R., Trotter, M., Buxton, B., & Holden, S. (2001). Drug design by machine learning: support vector machines for pharmaceutical data analysis. *Computers & Chemistry*, *26*(1), 5–14. https://doi.org/10.1016/S0097-8485(01)00094-8

Carrion, I., Aleman, J. L. F., & Toval, A. (2011). Assessing the HIPAA standard in practice: PHR privacy policies. *2011 Annual International Conference of the IEEE Engineering in Medicine and Biology Society*, 2380–2383. https://doi.org/10.1109/IEMBS.2011.6090664

Castro, M., & Liskov, B. (2002). Practical byzantine fault tolerance and proactive recovery. *ACM Transactions on Computer Systems*, *20*(4), 398–461. https://doi.org/10.1145/571637.571640

Coustasse, A., Arvidson, C., & Rutsohn, P. (2010). Pharmaceutical Counterfeiting and the RFID Technology Intervention. *Journal of Hospital Marketing & Public Relations*, *20*(2), 100–115. https://doi.org/10.1080/15390942.2010.493369

Dagher, G. G., Mohler, J., Milojkovic, M., & Marella, P. B. (2018). Ancile: Privacy-preserving framework for access control and interoperability of electronic health records using blockchain technology. *Sustainable Cities and Society*, *39*, 283–297. https://doi.org/10.1016/j.scs.2018.02.014

Dash, S., Shakyawar, S. K., Sharma, M., & Kaushik, S. (2019). Big data in healthcare: management, analysis and future prospects. *Journal of Big Data*, *6*(1), 1–25. https://doi.org/10.1186/s40537-019-0217-0

De Aguiar, E. J., Faiçal, B. S., Krishnamachari, B., & Ueyama, J. (2020). A Survey of

Blockchain-Based Strategies for Healthcare. *ACM Computing Surveys*, *53*(2), 1–27. https://doi.org/10.1145/3376915

Djenna, A., & Eddine Saidouni, D. (2018). Cyber Attacks Classification in IoT-Based-Healthcare Infrastructure. *2018 2nd Cyber Security in Networking Conference (CSNet)*, 1–4. https://doi.org/10.1109/CSNET.2018.8602974

Dogaru, D. I., & Dumitrache, I. (2017). Cyber security in healthcare networks. *2017 E-Health and Bioengineering Conference (EHB)*, 414–417. https://doi.org/10.1109/EHB.2017.7995449

Dyrda, L. (2020). *The 5 most significant cyberattacks in healthcare for 2020*. Becker's Hospital Review. https://www.beckershospitalreview.com/cybersecurity/the-5-most-significant-cyberattacks-in-healthcare-for-2020.html

Engelhardt, M. A. (2017). Hitching Healthcare to the Chain: An Introduction to Blockchain Technology in the Healthcare Sector. *Technology Innovation Management Review*, *7*(10), 22–34. https://doi.org/10.22215/timreview/1111

Esposito, C., De Santis, A., Tortora, G., Chang, H., & Choo, K.-K. R. (2018). Blockchain: A Panacea for Healthcare Cloud-Based Data Security and Privacy? *IEEE Cloud Computing*, *5*(1), 31–37. https://doi.org/10.1109/MCC.2018.011791712

Fernandez-Carames, T. M., & Fraga-Lamas, P. (2020). Towards Post-Quantum Blockchain: A Review on Blockchain Cryptography Resistant to Quantum Computing Attacks. *IEEE Access*, *8*, 21091–21116. https://doi.org/10.1109/ACCESS.2020.2968985

Filkins, B. (2014). *Health Care Cyberthreat Report: Widespread Compromises Detected, Compliance Nightmare on Horizon*. https://www.redwoodmednet.org/projects/events/20150731/docs/Norse-SANS-Healthcare-Cyberthreat-Report2014.pdf

Fugh-Berman, A., & Ahari, S. (2007). Following the Script: How Drug Reps Make Friends and Influence Doctors. *PLoS Medicine*, *4*(4), e150. https://doi.org/10.1371/journal.pmed.0040150

Gabison, G. (2016). Policy Considerations for the BlockchainTechnology Public and Private Applications. *Science and Technology Law Review*, *19*(3), 327–350. https://scholar.smu.edu/cgi/viewcontent.cgi?article=1043&context=scitech

Gahi, Y., Guennoun, M., & Mouftah, H. T. (2016). Big Data Analytics: Security and privacy challenges. *2016 IEEE Symposium on Computers and Communication (ISCC)*, 952–957. https://doi.org/10.1109/ISCC.2016.7543859

Gueron, S., Johnson, S., & Walker, J. (2011). SHA-512/256. *2011 Eighth International Conference on Information Technology: New Generations*, 354–358. https://doi.org/10.1109/ITNG.2011.69

Han, R., Foutris, N., & Kotselidis, C. (2019). Demystifying Crypto-Mining: Analysis and Optimizations of Memory-Hard PoW Algorithms. *2019 IEEE International Symposium on Performance Analysis of Systems and Software (ISPASS)*, 22–33. https://doi.org/10.1109/ISPASS.2019.00011

Hankerson, D., Menezes, A. J., & Vanstone, S. (2006). *Guide to Elliptic Curve Cryptography*. Springer-Verlag. https://doi.org/10.1007/b97644

Hao, X., Yu, L., Zhiqiang, L., Zhen, L., & Dawu, G. (2018). Dynamic Practical Byzantine Fault Tolerance. *2018 IEEE Conference on Communications and Network Security (CNS)*, 1–8. https://doi.org/10.1109/CNS.2018.8433150

Hardin, T., & Kotz, D. (2019). Blockchain in Health Data Systems: A Survey. *2019 Sixth International Conference on Internet of Things: Systems, Management and Security (IOTSMS)*, 490–497. https://doi.org/10.1109/IOTSMS48152.2019.8939174

Hasan, H. R., Salah, K., Jayaraman, R., Arshad, J., Yaqoob, I., Omar, M., & Ellahham, S. (2020). Blockchain-Based Solution for COVID-19 Digital Medical Passports and Immunity Certificates. *IEEE Access*, *8*, 222093–222108. https://doi.org/10.1109/ACCESS.2020.3043350

Hathaliya, J. J., & Tanwar, S. (2020). An exhaustive survey on security and privacy issues in Healthcare 4.0. *Computer Communications*, *153*, 311–335. https://doi.org/10.1016/j.comcom.2020.02.018

Hayrinen, K., Saranto, K., & Nykanen, P. (2008). Definition, structure, content, use and impacts of electronic health records: A review of the research literature. *International Journal of Medical Informatics*, *77*(5), 291–304. https://doi.org/10.1016/j.ijmedinf.2007.09.001

Heston, T. (2017). *A Case Study in Blockchain Healthcare Innovation* (AUTHOREA_213011_3643634).

Hölbl, M., Kompara, M., Kamišalić, A., & Nemec Zlatolas, L. (2018). A Systematic Review of the Use of Blockchain in Healthcare. *Symmetry*, *10*(10), 470. https://doi.org/10.3390/sym10100470

Hsiao, S.-C., & Kao, D.-Y. (2018). The static analysis of WannaCry ransomware. *2018 20th International Conference on Advanced Communication Technology (ICACT)*, 153–158. https://doi.org/10.23919/ICACT.2018.8323680

Huda, M. N., Sonehara, N., & Yamada, S. (2009). A Privacy Management Architecture For Patient-Controlled Personal Health Record System. *Journal of Engineering Science and Technology (JESTEC)*, *4*(2), 154–170.

Hussain, A., Heidemann, J., & Papadopoulos, C. (2003). A framework for classifying denial of service attacks. *Proceedings of the 2003 Conference on Applications, Technologies, Architectures, and Protocols for Computer Communications - SIGCOMM '03*, 99. https://doi.org/10.1145/863955.863968

Ibáñez, J. W., & Moccia, S. (2020). Designing the Architecture of a Blockchain Platform: The Case of Alastria, a National Public Permissioned Blockchain. *International Journal of Enterprise Information Systems*, *16*(3), 34–48. https://doi.org/10.4018/IJEIS.2020070103

Iroju, O., Soriyan, A., Gambo, I., J, & Olaleke. (2013). Interoperability in Healthcare: Benefits, Challenges and Resolutions. *International Journal of Innovation and Applied Studies*, *3*(1), 262–270.

Jaberidoost, M., Nikfar, S., Abdollahiasl, A., & Dinarvand, R. (2013). Pharmaceutical supply chain risks: a systematic review. *DARU Journal of Pharmaceutical Sciences*, *21*(1), 69. https://doi.org/10.1186/2008-2231-21-69

Kalla, A., Hewa, T., Mishra, R. A., Ylianttila, M., & Liyanage, M. (2020). The Role of Blockchain to Fight Against COVID-19. *IEEE Engineering Management Review*, *48*(3), 85–96. https://doi.org/10.1109/EMR.2020.3014052

Kawamoto, D. (2017). *IoT Security Incidents Rampant and Costly*. Dark Reading. https://www.darkreading.com/vulnerabilities---threats/iot-security-incidents-rampant-and-costly/d/d-id/1329367

Kehrli, J. (2016). *Blockchain 2.0-from bitcoin transactions to smart contract applications*. NiceIdeas. https://www.niceideas.ch/roller2/badtrash/entry/blockchain-2-0-from-bitcoin

Khezr, S., Moniruzzaman, M., Yassine, A., & Benlamri, R. (2019). Blockchain Technology in Healthcare: A Comprehensive Review and Directions for Future Research. *Applied Sciences*, *9*(9), 1736. https://doi.org/10.3390/app9091736

Koh, H. C., & Tan, G. (2005). Data mining applications in healthcare. *Journal of Healthcare Information Management : JHIM*, *19*(2), 64–72. http://www.ncbi.nlm.nih.gov/pubmed/15869215

Kramer, M. (2019). An Overview of Blockchain Technology Based on a Study of Public Awareness. *Global Journal of Business Research*, *13*(1), 83–91. https://ssrn.com/abstract=3381119

Krombholz, K., Hobel, H., Huber, M., & Weippl, E. (2015). Advanced social engineering attacks. *Journal of Information Security and Applications*, *22*, 113–122. https://doi.org/10.1016/j.jisa.2014.09.005

Kumar, V., & Walker, C. (2017). Cyber-Attacks: Rising Threat to Healthcare. *Vascular Disease Management*, *14*(3).

Landry, S., Beaulieu, M., & Roy, J. (2016). Strategy deployment in healthcare services: A case study approach. *Technological Forecasting and Social Change*, *113*, 429–437. https://doi.org/10.1016/j.techfore.2016.09.006

Le, T. T., Andreadakis, Z., Kumar, A., Gómez Román, R., Tollefsen, S., Saville, M., & Mayhew, S. (2020). The COVID-19 vaccine development landscape. *Nature Reviews Drug*

*Discovery*, *19*(5), 305–306. https://doi.org/10.1038/d41573-020-00073-5

Lee, C. C. M., Thampi, S., Lewin, B., Lim, T. J. D., Rippin, B., Wong, W. H., & Agrawal, R. V. (2020). Battling COVID-19: critical care and peri-operative healthcare resource management strategies in a tertiary academic medical centre in Singapore. *Anaesthesia*, *75*(7), 861–871. https://doi.org/10.1111/anae.15074

Li, P., Xu, C., Jin, H., Hu, C., Luo, Y., Cao, Y., Mathew, J., & Ma, Y. (2020). ChainSDI: A Software-Defined Infrastructure for Regulation-Compliant Home-Based Healthcare Services Secured by Blockchains. *IEEE Systems Journal*, *14*(2), 2042–2053. https://doi.org/10.1109/JSYST.2019.2937930

Liu, X., Wang, Z., Jin, C., Li, F., & Li, G. (2019). A Blockchain-Based Medical Data Sharing and Protection Scheme. *IEEE Access*, *7*, 118943–118953. https://doi.org/10.1109/ACCESS.2019.2937685

Madine, M. M., Battah, A. A., Yaqoob, I., Salah, K., Jayaraman, R., Al-Hammadi, Y., Pesic, S., & Ellahham, S. (2020). Blockchain for Giving Patients Control Over Their Medical Records. *IEEE Access*, *8*, 193102–193115. https://doi.org/10.1109/ACCESS.2020.3032553

Mandl, K. D., & Kohane, I. S. (2012). Escaping the EHR Trap — The Future of Health IT. *New England Journal of Medicine*, *366*(24), 2240–2242. https://doi.org/10.1056/NEJMp1203102

Meena, D. K., Dwivedi, R., & Shukla, S. (2019). Preserving Patient's Privacy using Proxy Re-encryption in Permissioned Blockchain. *2019 Sixth International Conference on Internet of Things: Systems, Management and Security (IOTSMS)*, 450–457. https://doi.org/10.1109/IOTSMS48152.2019.8939226

Meng, W., Li, W., & Zhu, L. (2020). Enhancing Medical Smartphone Networks via Blockchain-Based Trust Management Against Insider Attacks. *IEEE Transactions on Engineering Management*, *67*(4), 1377–1386. https://doi.org/10.1109/TEM.2019.2921736

Moulis, G., Lapeyre-Mestre, M., Palmaro, A., Pugnet, G., Montastruc, J.-L., & Sailler, L. (2015). French health insurance databases: What interest for medical research? *La Revue de Médecine Interne*, *36*(6), 411–417. https://doi.org/10.1016/j.revmed.2014.11.009

Mullard, A. (2020). COVID-19 vaccine development pipeline gears up. *The Lancet*, *395*(10239), 1751–1752. https://doi.org/10.1016/S0140-6736(20)31252-6

Muthuppalaniappan, M., & Stevenson, K. (2021). Healthcare cyber-attacks and the COVID-19 pandemic: an urgent threat to global health. *International Journal for Quality in Health Care*, *33*(1). https://doi.org/10.1093/intqhc/mzaa117

Nakamato, S. (2008). *Bitcoin: A Peer-to-Peer Electronic Cash System*. https://bitcoin.org/bitcoin.pdf

Nawari, N. O., & Ravindran, S. (2019). Blockchain and the built environment: Potentials and limitations. *Journal of Building Engineering*, *25*, 100832 (1-16).

https://doi.org/10.1016/j.jobe.2019.100832

Nguyen, D. C., Pathirana, P. N., Ding, M., & Seneviratne, A. (2019). Blockchain for Secure EHRs Sharing of Mobile Cloud Based E-Health Systems. *IEEE Access*, *7*, 66792–66806. https://doi.org/10.1109/ACCESS.2019.2917555

Ongaro, D.; Ousterhout, J. (2014). In Search of an Understandable Consensus Algorithm. *Proceedings of USENIX ATC '14: 2014 USENIX Annual Technical Conference.*, 305–319. https://www.usenix.org/system/files/conference/atc14/atc14-paper-ongaro.pdf

Pandey, A. K., Khan, A. I., Abushark, Y. B., Alam, M. M., Agrawal, A., Kumar, R., & Khan, R. A. (2020). Key Issues in Healthcare Data Integrity: Analysis and Recommendations. *IEEE Access*, *8*, 40612–40628. https://doi.org/10.1109/ACCESS.2020.2976687

Partala, J., Nguyen, T. H., & Pirttikangas, S. (2020). Non-Interactive Zero-Knowledge for Blockchain: A Survey. *IEEE Access*, *8*, 227945–227961. https://doi.org/10.1109/ACCESS.2020.3046025

Pitacco, E. (2014). *Health Insurance*. Springer International Publishing. https://doi.org/10.1007/978-3-319-12235-9

Plachkinova, M., Andres, S., & Chatterjee, S. (2015). A Taxonomy of mHealth Apps -- Security and Privacy Concerns. *2015 48th Hawaii International Conference on System Sciences*, 3187–3196. https://doi.org/10.1109/HICSS.2015.385

Qiao, R., Luo, X.-Y., Zhu, S.-F., Liu, A.-D., Yan, X.-Q., & Wang, Q.-X. (2020). Dynamic Autonomous Cross Consortium Chain Mechanism in e-Healthcare. *IEEE Journal of Biomedical and Health Informatics*, *24*(8), 2157–2168. https://doi.org/10.1109/JBHI.2019.2963437

Radanović, I., & Likić, R. (2018). Opportunities for Use of Blockchain Technology in Medicine. *Applied Health Economics and Health Policy*, *16*(5), 583–590. https://doi.org/10.1007/s40258-018-0412-8

Rajput, A. R., Li, Q., Taleby Ahvanooey, M., & Masood, I. (2019). EACMS: Emergency Access Control Management System for Personal Health Record Based on Blockchain. *IEEE Access*, *7*, 84304–84317. https://doi.org/10.1109/ACCESS.2019.2917976

Raza, M., Iqbal, M., Sharif, M., & Haider, W. (2012). A survey of password attacks and comparative analysis on methods for secure authentication. *World Applied Sciences Journal*, *19*(4), 439–444. https://doi.org/10.5829/idosi.wasj.2012.19.04.1837

Ruiz, J. (2020). *Public-Permissioned blockchains as Common-Pool Resources*. Linkedin. https://www.linkedin.com/pulse/public-permissioned-blockchains-common-pool-resources-jesus-ruiz/

Saldamli, G., Reddy, V., Bojja, K. S., Gururaja, M. K., Doddaveerappa, Y., & Tawalbeh, L. (2020). Health Care Insurance Fraud Detection Using Blockchain. *2020 Seventh*

*International Conference on Software Defined Systems (SDS)*, 145–152. https://doi.org/10.1109/SDS49854.2020.9143900

Saleh, F. (2021). Blockchain without Waste: Proof-of-Stake. *The Review of Financial Studies*, *34*(3), 1156–1190. https://doi.org/10.1093/rfs/hhaa075

Seh, A. H., Zarour, M., Alenezi, M., Sarkar, A. K., Agrawal, A., Kumar, R., & Ahmad Khan, R. (2020). Healthcare Data Breaches: Insights and Implications. *Healthcare*, *8*(2), 133. https://doi.org/10.3390/healthcare8020133

Shahnaz, A., Qamar, U., & Khalid, A. (2019). Using Blockchain for Electronic Health Records. *IEEE Access*, *7*, 147782–147795. https://doi.org/10.1109/ACCESS.2019.2946373

Shen, M., Duan, J., Zhu, L., Zhang, J., Du, X., & Guizani, M. (2020). Blockchain-Based Incentives for Secure and Collaborative Data Sharing in Multiple Clouds. *IEEE Journal on Selected Areas in Communications*, *38*(6), 1229–1241. https://doi.org/10.1109/JSAC.2020.2986619

Shi, S., He, D., Li, L., Kumar, N., Khan, M. K., & Choo, K.-K. R. (2020). Applications of blockchain in ensuring the security and privacy of electronic health record systems: A survey. *Computers & Security*, *97*, 101966. https://doi.org/10.1016/j.cose.2020.101966

Smith, B., Xiong, J., & Medlin, D. (2020). Case Study of Blockchain Applications in Supply Chain Management- Opportunities andChallenges. *2020 Proceedings of the Conference on Information Systems Applied Research Virtual Conference*, 1501–1508.

Stafford, T. F., & Treiblmaier, H. (2020). Characteristics of a Blockchain Ecosystem for Secure and Sharable Electronic Medical Records. *IEEE Transactions on Engineering Management*, *67*(4), 1340–1362. https://doi.org/10.1109/TEM.2020.2973095

Stephen, O., Sain, M., Maduh, U. J., & Jeong, D.-U. (2019). An Efficient Deep Learning Approach to Pneumonia Classification in Healthcare. *Journal of Healthcare Engineering*, *2019*, 1–7. https://doi.org/10.1155/2019/4180949

Sun, Y. L., Wei, Y., Zhu, H., & Liu, K. J. R. (2006). Information theoretic framework of trust modeling and evaluation for ad hoc networks. *IEEE Journal on Selected Areas in Communications*, *24*(2), 305–317. https://doi.org/10.1109/JSAC.2005.861389

Sunyaev, A., Chornyi, D., Mauro, C., & Krcmar, H. (2010). Evaluation Framework for Personal Health Records: Microsoft HealthVault Vs. Google Health. *2010 43rd Hawaii International Conference on System Sciences*, 1–10. https://doi.org/10.1109/HICSS.2010.192

Tariq, N., Qamar, A., Asim, M., & Khan, F. A. (2020). Blockchain and Smart Healthcare Security: A Survey. *Procedia Computer Science*, *175*, 615–620. https://doi.org/10.1016/j.procs.2020.07.089

Teklehaimanot, H. D., & Teklehaimanot, A. (2013). Human resource development for a community-based health extension program: a case study from Ethiopia. *Human Resources*

*for Health*, *11*(1), 39. https://doi.org/10.1186/1478-4491-11-39

Tomaz, A. E. B., Nascimento, J. C. Do, Hafid, A. S., & De Souza, J. N. (2020). Preserving Privacy in Mobile Health Systems Using Non-Interactive Zero-Knowledge Proof and Blockchain. *IEEE Access*, *8*, 204441–204458. https://doi.org/10.1109/ACCESS.2020.3036811

Ulieru, M. (2016). Blockchain 2.0 and Beyond: Adhocracies. In *Banking Beyond Banks and Money* (pp. 297–303). Springer, Cham. https://doi.org/10.1007/978-3-319-42448-4_15

Urciuoli, L., Männistö, T., Hintsa, J., & Khan, T. (2013). Supply Chain Cyber Security – Potential Threats. *Information & Security: An International Journal*, *29*, 51–68. https://doi.org/10.11610/isij.2904

Wang, S., Zhang, D., & Zhang, Y. (2019). Blockchain-Based Personal Health Records Sharing Scheme With Data Integrity Verifiable. *IEEE Access*, *7*, 102887–102901. https://doi.org/10.1109/ACCESS.2019.2931531

Wang, Y., Zhang, A., Zhang, P., & Wang, H. (2019). Cloud-Assisted EHR Sharing With Security and Privacy Preservation via Consortium Blockchain. *IEEE Access*, *7*, 136704–136719. https://doi.org/10.1109/ACCESS.2019.2943153

Win, K. T., Susilo, W., & Mu, Y. (2006). Personal Health Record Systems and Their Security Protection. *Journal of Medical Systems*, *30*(4), 309–315. https://doi.org/10.1007/s10916-006-9019-y

Wright, T. (2020). *Blockchain App Used to Track COVID-19 Cases in Latin America*. Coin Telegraph: The Future of Money. https://cointelegraph.com/news/blockchain-app-used-to-track-covid-19-cases-in-latin-america

Wust, K., & Gervais, A. (2018). Do you Need a Blockchain? *2018 Crypto Valley Conference on Blockchain Technology (CVCBT)*, 45–54. https://doi.org/10.1109/CVCBT.2018.00011

Xia, Q., Sifah, E., Smahi, A., Amofa, S., & Zhang, X. (2017). BBDS: Blockchain-Based Data Sharing for Electronic Medical Records in Cloud Environments. *Information*, *8*(2), 44. https://doi.org/10.3390/info8020044

Xiao, Y., Zhang, N., Lou, W., & Hou, Y. T. (2020). A Survey of Distributed Consensus Protocols for Blockchain Networks. *IEEE Communications Surveys & Tutorials*, *22*(2), 1432–1465. https://doi.org/10.1109/COMST.2020.2969706

Xu, H., Zhang, L., Onireti, O., Fang, Y., Buchanan, W. J., & Imran, M. A. (2021). BeepTrace: Blockchain-Enabled Privacy-Preserving Contact Tracing for COVID-19 Pandemic and Beyond. *IEEE Internet of Things Journal*, *8*(5), 3915–3929. https://doi.org/10.1109/JIOT.2020.3025953

Xu, J., Xue, K., Li, S., Tian, H., Hong, J., Hong, P., & Yu, N. (2019). Healthchain: A Blockchain-Based Privacy Preserving Scheme for Large-Scale Health Data. *IEEE Internet*

*of Things Journal*, *6*(5), 8770–8781. https://doi.org/10.1109/JIOT.2019.2923525

Yazdinejad, A., Srivastava, G., Parizi, R. M., Dehghantanha, A., Choo, K.-K. R., & Aledhari, M. (2020). Decentralized Authentication of Distributed Patients in Hospital Networks Using Blockchain. *IEEE Journal of Biomedical and Health Informatics*, *24*(8), 2146–2156. https://doi.org/10.1109/JBHI.2020.2969648

Yeoh, P. (2017). Regulatory issues in blockchain technology. *Journal of Financial Regulation and Compliance*, *25*(2), 196–208. https://doi.org/10.1108/JFRC-08-2016-0068

Yli-Huumo, J., Ko, D., Choi, S., Park, S., & Smolander, K. (2016). Where is current research on Blockchain technology? - A systematic review. *PLoS ONE*, *11*(10), 1–27. https://doi.org/10.1371/journal.pone.0163477

Yu, K., Tan, L., Shang, X., Huang, J., Srivastava, G., & Chatterjee, P. (2021). Efficient and Privacy-Preserving Medical Research Support Platform Against COVID-19: A Blockchain-Based Approach. *IEEE Consumer Electronics Magazine*, *10*(2), 111–120. https://doi.org/10.1109/MCE.2020.3035520

Zghaibeh, M., Farooq, U., Hasan, N. U., & Baig, I. (2020). SHealth: A Blockchain-Based Health System With Smart Contracts Capabilities. *IEEE Access*, *8*, 70030–70043. https://doi.org/10.1109/ACCESS.2020.2986789

Zhang, A., & Lin, X. (2018). Towards Secure and Privacy-Preserving Data Sharing in e-Health Systems via Consortium Blockchain. *Journal of Medical Systems*, *42*(8), 1–18. https://doi.org/10.1007/s10916-018-0995-5

Zhang, S., & Lee, J.-H. (2020). Analysis of the main consensus protocols of blockchain. *ICT Express*, *6*(2), 93–97. https://doi.org/10.1016/j.icte.2019.08.001

Zhu, P., Hu, J., Zhang, Y., & Li, X. (2020). A Blockchain Based Solution for Medication Anti-Counterfeiting and Traceability. *IEEE Access*, *8*, 184256–184272. https://doi.org/10.1109/ACCESS.2020.3029196

Zhuang, Y., Sheets, L. R., Chen, Y.-W., Shae, Z.-Y., Tsai, J. J. P., & Shyu, C.-R. (2020). A Patient-Centric Health Information Exchange Framework Using Blockchain Technology. *IEEE Journal of Biomedical and Health Informatics*, *24*(8), 2169–2176. https://doi.org/10.1109/JBHI.2020.2993072

Zou, W., Lo, D., Kochhar, P. S., Le, X.-B. D., Xia, X., Feng, Y., Chen, Z., & Xu, B. (2019). Smart Contract Development: Challenges and Opportunities. *IEEE Transactions on Software Engineering*, 1–1. https://doi.org/10.1109/TSE.2019.2942301